\documentclass[pdflatex,sn-mathphys-ay]{sn-jnl}

\usepackage{color}
\usepackage{setspace}
\usepackage{caption} 
\usepackage{graphicx}
\usepackage{url}
\usepackage{hyperref}
\usepackage{graphicx}%
\usepackage{multirow}%
\usepackage{amsmath,amssymb,amsfonts}%
\usepackage{amsthm}%
\usepackage{mathrsfs}%
\usepackage[title]{appendix}%
\usepackage{textcomp}%
\usepackage{manyfoot}%
\usepackage{booktabs}%
\usepackage{algorithm}%
\usepackage{algorithmicx}%
\usepackage{algpseudocode}%
\usepackage{listings}%
\usepackage{comment}
\usepackage{xcolor}

\theoremstyle{thmstyleone}%
\theoremstyle{thmstyletwo}%

\theoremstyle{thmstylethree}%

\begin{document}


\title{A Conceptual Model and Methodology for Sustainability-aware, IoT-enhanced Business Processes} 


\author*[1]{\fnm{Victoria} \sur{Torres}}\email{victoria.torres@upv.es}
\equalcont{These authors contributed equally to this work.}

\author[2]{\fnm{Ronny} \sur{Seiger}}\email{ronny.seiger@rwth-aachen.de}
\equalcont{These authors contributed equally to this work.}

\author[1]{\fnm{Manoli} \sur{Albert}}\email{malbert@dsic.upv.es}

\author[1]{\fnm{Antoni} \sur{Mestre}}\email{anmesgas@vrain.upv.es}

\author[1]{\fnm{Pedro} \sur{Valderas}}\email{pvalderas@dsic.upv.es}

\affil[1]{\orgdiv{VRAIN-PROS}, \orgname{Universitat Polit\`ecnica de Val\`encia}, \city{Valencia}, \country{Spain}}

\affil[2]{\orgdiv{Faculty of Computer Science}, \orgname{RWTH Aachen University}, \city{Aachen}, \country{Germany}}



\abstract{
The real-time data collection and automation capabilities offered by the Internet of Things (IoT) are revolutionizing and transforming Business Processes (BPs) into \emph{IoT-enhanced BPs}, showing high potential for improving sustainability. Although already studied in Business Process Management (BPM), sustainability research has primarily focused on environmental concerns. However, achieving a holistic and lasting impact requires a systematic approach to address sustainability beyond the environmental dimension. This work proposes a conceptual model and a structured methodology with the goal of analyzing the potential of IoT to measure and improve the sustainability of BPs. The conceptual model formally represents key sustainability concepts, linking BPM and IoT by highlighting how IoT devices support and contribute to sustainability. The methodology guides the systematic analysis of existing BPs, identifies opportunities, and implements sustainability-aware, IoT-enhanced BPs. The approach is illustrated through a running example from the tourism domain and a controlled case study in healthcare.
}


\maketitle

\section{Introduction}
In today's interconnected world, businesses are increasingly integrating Internet of Things (IoT) devices into their operations. Several works have acknowledged the important roles that IoT devices can play in the context of Business Process Management (BPM)~\citep{janiesch2020internet} to streamline their operations, optimize resource utilization, and improve overall productivity. While some IoT devices (actuators) can act as active process performers that are able to support and execute automated process activities~\citep{seiger2022integrating,seiger2018case}, others (sensors) can monitor the environment, the process executions, and detect changes that may trigger events in a business process (BP)~\citep{DBLP:journals/jss/ValderasTS22,kirikkayis2023integrating}.

This symbiotic relationship between IoT devices and BPs, hereinafter \emph{IoT-enhanced BPs}~\citep{DBLP:journals/jss/ValderasTS22}, has revolutionized the way organizations operate, offering unprecedented insights and capabilities, also in terms of sustainability, where IoT devices can be employed to improve sustainability across various dimensions (e.g., by enabling real-time monitoring of energy consumption to reduce resource waste). While concern for sustainability in BPs began to take shape in the 1970s and 1980s, it has been over the past two decades that it has been more deeply and strategically integrated into the BPM field. For instance, a lot of research efforts have been made from the area of Green BPM~\citep{DBLP:journals/ajis/GhoseHHL09} to reduce the environmental impact of BPs. However, sustainability is not limited only to the environmental dimension. It is a broader notion and should be extended to other dimensions as identified in~\cite{DBLP:conf/aosd/PenzenstadlerF13}. More specifically, sustainability encompasses economic, social, human, environmental, and technical aspects, all of which are interconnected and are essential for long-term viability. Therefore, analyzing and improving the sustainability of BPs--with and without IoT enhancements--from these five different dimensions becomes paramount for organizations striving to align with sustainability goals and mitigate potential risks.

In the context of growing environmental concerns, regulatory pressures, and corporate social responsibility commitments, organizations are increasingly expected to ensure that their BPs contribute positively to sustainability goals. However, despite this growing relevance, many companies still lack a systematic and integrated approach to assess and improve their sustainability performance. Existing frameworks and methodologies often focus on operational efficiency or economic outcomes, but rarely address sustainability in a comprehensive and measurable manner. Without robust conceptual foundations and clear methodological guidelines, it becomes challenging to identify sustainability gaps, implement improvements, and monitor their effectiveness over time. Therefore, there is a clear need to develop a conceptual framework and a practical methodology that allow organizations to systematically measure, analyze, and improve their BPs from a sustainability perspective, ensuring that sustainability is embedded not only in strategic goals but also in day-to-day operations. 

To address these aspects, we propose a conceptual model in the form of a well-defined metamodel in combination with a structured methodology. The metamodel provides the necessary abstraction and formalism to represent process elements and sustainability-related attributes in a consistent and extensible manner and it bridges the gap to the IoT domain to show application opportunities for these new technologies. Here, IoT can support sustainability-related improvements in existing BPs in various ways. For example, sensors combined with BP data can be used to measure and analyze the sustainability performance of specific BP parts. Additionally, IoT devices can support the automation of tasks aimed at improving the overall sustainability of BPs. In addition, the methodology offers a guided approach for instantiating the metamodel in real-world scenarios, enabling practitioners to systematically measure and improve BPs through IoT technologies. 

\subsection{Problem statement and research question} \label{sec:problemStatement}

Despite the many new opportunities offered by the integration of BPM and IoT~\citep{janiesch2020internet}, and the increasing importance of sustainability considerations in BPs~\citep{DBLP:journals/ajis/GhoseHHL09}, there is a lack of a holistic approach that systematically considers the integration of IoT technologies to support, analyze and improve sustainability-related effects in BPs. In this paper, we adopt the basic assumption that we do not aim to identify or design entirely new BPs. Instead, we start from existing processes, analyze them to uncover their improvement potential regarding sustainability, explore how IoT can support these improvements, and then assess the impact of the proposed changes. To this end, the problem addressed in this paper can be stated through the following research question (RQ): \textbf{How can we provide methodological support to (1)~analyze and, (2)~improve the sustainability of Business Processes by using IoT?}

\subsection{Main objectives and contributions} \label{sec:contributions}

The main objective of this work is to answer the above-introduced research question by providing a solution to face the incorporation of the five sustainability dimensions into the design, analysis, and improvement of BPs with IoT enhancements. 
We aim to take the concept of IoT-enhanced BPs one step further to achieve \emph{Sustainability-aware, IoT-enhanced BPs}. To do so, the contributions of this paper are the following:

 \begin{enumerate}

    \item \textbf{Creation of a conceptual model}. We need to precisely characterize how the notion of sustainability and its five dimensions correlates with concepts from the fields of BPM and IoT. To do so, we propose a conceptual model in the form of a metamodel based on the work of~\cite{DBLP:conf/aosd/PenzenstadlerF13}, defining the essential concepts and their relationships to bridge the fields of sustainability, BPM and IoT, and thus to address sustainability in IoT-enhanced BPs.
    \item \textbf{Definition of a methodology}. We define how this conceptual model can be used in the analysis and development of Sustainability-aware, IoT-enhanced BPs. To achieve this goal, we propose a methodology inspired by the work of~\cite{larsch2017integrating} to systemically identify, measure, and improve sustainability-related aspects in BPs via IoT technologies.

 \end{enumerate}

 \subsection{Research method and paper structure}

 A methodology in line with the precepts of Design Science Research (DSR)~\citep{hevner2004design,peffers2007design} is used for this research work. DSR aims at developing practical solutions that can be used by professionals in the field of information systems engineering. More concretely, solutions, or design artifacts, can take the form of constructs, models, methods, or instantiations~\citep{hevner2004design}. Considering the contributions presented above, the artifacts we have developed are the conceptual model and a methodology to develop and analyze sustainability-aware, IoT-enhanced BPs. According
to~\cite{hevner2004design}, these artifacts can be classified as a model and a method, respectively, since they are used to define a set of actionable instructions that are conceptual and not algorithmic. 

We applied the DSR methodology (DSRM) to develop these artifacts and performed the six activities proposed by~\cite{peffers2007design} following a problem-centered approach. These six activities are: (1)~Problem identification and motivation; (2)~Define the objectives for a solution; (3)~Design and development; (4)~Demonstration; (5)~Evaluation; and (6)~Communication. 
We first identified the specific research problem and motivated it, which was presented at the beginning of this section. This motivation is complemented by the study of relevant background concepts in Section~\ref{sec:background}, and of the state of the art that is presented in Section~\ref{sec:related}, which identifies shortcomings of related approaches with respect to our main objectives. Next, we defined the objectives of the solution, which are presented in Sections~\ref{sec:problemStatement} and~\ref{sec:contributions}. The next activity in the DSRM consists of the design and development of the artifacts required to support the proposed objectives. To do so, we followed an action-research development~\citep{avison1999action} to iteratively study the problem to solve, apply some actions, and analyze if the obtained results satisfy our purposes. This is explained in Sections~\ref{sec:framework} and~\ref{sec:methodology}, which introduce the two main artifacts--the conceptual model and the methodology, respectively. To explore the key concepts in a controlled and simplified environment, these artifacts are instantiated and illustrated with a BP--verified by a domain expert--representing the management of the stay of a hotel guest, as presented in Section~\ref{sub:runningexample}. The next activities to be performed are demonstration and validation, in which the developed artifacts are used to solve one or more instances of the considered problem and it is analyzed how well the artifacts support a solution to the problems. Section~\ref{sec:case-studies} presents a controlled case study of a real-world case from the domain of smart healthcare, which was developed together with medical professionals. The study focuses on the monitoring and tracking of medical procedures through the pragmatical use of the proposed method to evaluate and improve BPs in smart healthcare through IoT. Section~\ref{sec:discussion} discusses the results of our work to analyze the suitability of the proposed artifacts. To complete the DSRM, we are communicating our results to the research community through this paper, whose last section presents some conclusions and provides insights into directions for future work.

\section{Scenario and background} 
\label{sec:background}

This section provides the context necessary for understanding this work. It begins by introducing a real-world BP that will be used throughout the work as a running example. Then, the section explores the \emph{Sustainability} concept with a particular focus on its contextualization within the BPM lifecycle. Then, we describe the notion of \emph{IoT-enhanced Business Processes}, examining how the integration of IoT devices into business operations can transform efficiency, decision making, and overall productivity.

\subsection{The hotel stay process example}
\label{sub:runningexample}
\begin{figure}
    \centering
    \includegraphics[width=1\linewidth]{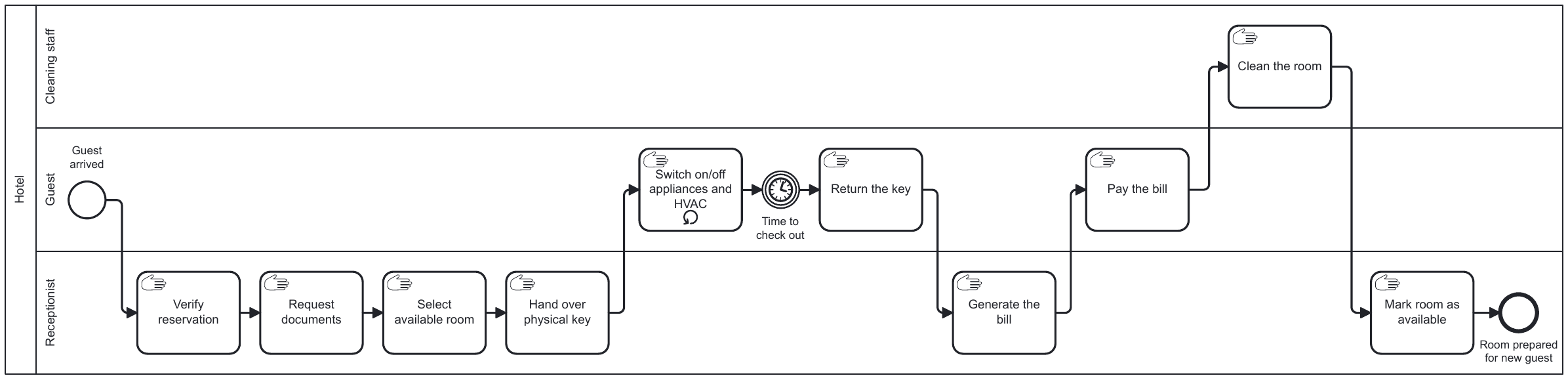}
    \caption{Process model (in BPMN 2.0) illustrating the stay of a guest in a hotel}
    \label{fig:check-in-process-example}
\end{figure}

This section describes a fully manual process, with no automation for managing the stay of a guest in a small hotel (cf. Fig.~\ref{fig:check-in-process-example}). The process involves the processing of the guest upon its arrival, including the verification of the reservation and the guest's documents, selecting an available room, and the handing over of the physical key. During the stay, the guest usually operates technical appliances, lights and HVAC (heating, ventilation, air conditioning) manually. Upon checkout, the key is returned by the guest, the bill generated by the receptionist and paid by the guest. Finally, the cleaning staff prepares the room for the next guest, and the receptionist then marks it as being available. 

The relevance of this process to sustainability lies in the opportunities for improvements in different aspects, e.g., 1) the manual operation of energy appliances could be replaced by smart systems to optimize energy consumption, and 2) the digitization of documentation and billing could reduce paper usage and streamline guest handling. This process serves as running example to illustrate the relevant concepts and steps in the proposed methodology. The example and its application following the methodology have been reviewed by an expert working on sustainability-related aspects in a management position at a hotel in Spain.

\subsection{Sustainability}
\label{sec:sustainability}
Although sustainability, in its various dimensions, is now deeply embedded in our society, it is by no means a new topic of discussion. In fact, it has been on the global agenda for several decades. A clear milestone is the Brundtland Report~\citep{keeble1988brundtland}, published in 1987, which defined \emph{sustainable development} as ``development that meets the needs of the present without compromising the ability of future generations to meet their own needs.'' Besides this definition, we find other references that collectively offer a foundational understanding of sustainability. These include~\cite{elkington1994cannibals} that introduced the concept of the \emph{Triple Bottom Line} which expanded the understanding of sustainability by emphasizing the importance of balancing environmental, social, and economic factors, often summarized as ``People, Planet, Profit''. The authors in~\cite{kates2005sustainable} outlined the goals, indicators, and values of sustainable development, providing a detailed framework for understanding and implementing sustainability across different sectors and disciplines. With~\cite{unitednations2015transforming} the United Nations provided a comprehensive blueprint for achieving sustainability globally, addressing a wide range of issues from poverty and inequality to climate change and environmental degradation.

Although the essence of sustainability in all these references remains constant in terms of its goal to balance present and future needs, its definition and application can vary depending on the specific context where it is applied to better accommodate different disciplines. However, all disciplines share the idea that sustainability consists of various \emph{dimensions}, some of which are considered relevant by all (i.e., the social, economic, and environmental dimensions) and others are more specific to the discipline in question (e.g., the governance, technological, cultural, or knowledge dimensions)~\citep{DBLP:journals/bpmj/CouckuytL20}.

Within the Software Engineering field,~\cite{Karlskrona-Manifesto} and~\cite{VENTERS2023107316} highlight that sustainability design requires cross-disciplinary expertise covering not just different dimensions but most importantly, the interaction between these and additional aspects. In the context of IoT-enhanced BPs, which this work focuses on, it encompasses the integration of IoT technologies into various aspects of business operations to drive efficiency, innovation, and competitive advantage. While the sustainability model in~\cite{DBLP:conf/aosd/PenzenstadlerF13} provides a comprehensive basis for analysis and potential improvements of sustainability aspects, it does not link to the field of BPM and any specific activity domains, nor does it allow the explicit representation and discussion of IoT devices and technologies in the context of BPs.

\subsection{IoT-enhanced Business Processes} \label{sec:iot-bp}

Several works have acknowledged the important roles that IoT devices can play in the context of BPM~\citep{janiesch2020internet}. Specifically, IoT actuators might act as active process performers that are able to support and execute automated process activities that affect the environment by executing an action (e.g.,~opening a door, moving an object)~\citep{seiger2022integrating,seiger2018case}. Sensors, in turn, monitor the environment and detect changes that may trigger events in a BP~\citep{DBLP:journals/jss/ValderasTS22,kirikkayis2023integrating}. Moreover, sensors can act as passive participants monitoring a)~the execution of process activities~\citep{franceschetti2023event,seiger2025online}, b)~interactions with process resources (e.g., humans or objects) in the physical world~\citep{bauer2013iot}, and c)~environmental (context) factors~\citep{bertrand2021bridging,bauer2013iot}. For example, in manufacturing processes, IoT-enabled machinery can autonomously adjust settings to optimize output while minimizing energy consumption~\citep{mangler2024internet}. In retail, IoT sensors can track inventory levels in real-time, enabling timely replenishment and reducing shortage. In logistics and supply chain management, IoT devices can provide accurate location tracking, temperature monitoring for perishable goods, and predictive maintenance for vehicles and equipment.

In the context of our work, we investigate sustainability aspects related to \emph{IoT-enhanced BPs}~\citep{DBLP:journals/jss/ValderasTS22}. Here, the IoT provides novel opportunities to directly measure process-related sustainability indicators via sensors. Using the IoT to support the automated execution of process activities might also positively contribute to sustainability in all dimensions~\citep{albert2024sustainability}. On the downside, introducing IoT devices to BPs for automation and monitoring purposes might also have a negative impact on sustainability as the new devices generate costs along the sustainability dimensions~\citep{becker2016requirements}. Among others, these costs refer to the provisioning, operation and maintenance of the IoT devices as well as to the necessary software and data processing infrastructure~\citep{gray2018energy}. 
Therefore, IoT devices themselves do have an impact on sustainability, both positive and negative. While they offer numerous benefits for measuring and improving sustainability in BPs, they also introduce certain sustainability challenges across different dimensions. Thus, considering sustainability in and through IoT-enhanced BPs necessitates an extensive analysis and trade-off discussion to find feasible solutions~\citep{lago2017reality}.

\section{Related work} \label{sec:related}

This section reviews the existing literature on the intersection of BPM and sustainability, i.e., how sustainability is addressed within the BPM field. By doing so, we aim to identify current trends, challenges and shortcomings in integrating sustainable practices within BPM frameworks. 

\subsection{Addressing sustainability in BPM}

Sustainability began to be addressed within the field of BPM from mainly an environmental perspective, aiming to measure and reduce the use of resources that negatively impact the environment. Within this context the term \textit{Green BPM} was used in the literature~\citep{DBLP:journals/ajis/GhoseHHL09,DBLP:conf/hicss/OpitzKK14,jakobi2016towards,kuppusamy2015green,lubbecke2015simulation,nowak2011differences,Seidel2012,vomBrocke2012}) to encompass works that face sustainability in BPM from different perspectives or with different purposes~\citep{DBLP:journals/bpmj/CouckuytL20}. Here, related works vary with respect to the goals sought: while some of them propose specific goals such as the reduction of carbon emissions~\citep{DBLP:journals/ajis/GhoseHHL09} or power consumption~\citep{nowak2013determining}, others address more general issues by considering various functional affordances of information systems required for sustainability transformations~\citep{Seidel2012}. Also, we can find approaches with different scopes since some just address one phase of the BPM lifecycle~\citep{schoormann2017sustainability} while others focus on all phases~\citep{nowak2011differences,larsch2017integrating}.  

Another term commonly used in this context is \textit{Sustainable BPM}. While some authors use it synonymously with \textit{Green BPM}~\citep{DBLP:journals/bpmj/CouckuytL20}, other works~\citep{ahmed2012sustainability,rozman2015achieving} use this notion to address sustainability more broadly, encompassing not only environmental aspects but also economic and social dimensions. Other works also provide a vision of sustainability that goes beyond the environmental perspective. For instance, Betz et al.~state that sustainability can have three levels of impacts on BPs: (1)~direct, which includes any impact directly caused by the production and use of software systems such as energy usage, e-waste production, emissions, etc.; (2)~indirect, which includes actions that are enabled by using software systems such as process improvements and behavioral or technological changes; and (3)~socioeconomic, which refers to changes on economic structures and on institutional level~\citep{betz2014sustainability}. In the same way, Becker et al.~advocate a systemic and interdisciplinary notion of sustainability with five key dimensions: economic, social, environmental, technical, and individual~\citep{becker2016requirements}.

In addition, different semantic characterizations based on metamodels have been proposed to conceptually integrate sustainability and BPM~\citep{larsch2017integrating,DBLP:conf/aosd/PenzenstadlerF13,bogatinovska2025metamodel}. However, none of them consider the use of IoT devices to measure sustainability along the different dimensions, improve it, or evaluate their impact on the sustainability of BPs as we aim to do. Other works focus on providing semantic descriptions of sustainability through ontologies to investigate general software systems instead of emphasizing on BPs~\citep{restrepo2024sinso}. In the same line, other works characterize different dimensions of sustainability from a general software perspective. For instance, the authors in~\cite{khalifeh2020incorporating} associate eight quality characteristics of the ISO/IEC 25010 product quality model with the environmental, economic, and social dimensions. The work~\cite{condori2019towards} uses different case studies to identify relevant quality attributes in the economic, technical, environmental, and social dimensions. Moreira et al.~propose a catalog of reusable sustainability requirements focused on the social and technical dimensions~\citep{moreira2023social}. Finally, several works identified attributes associated with the environmental dimension~\citep{kern2018sustainable,kocak2019utility,koccak2015integrating,garcia2021energy}. To complement these works, a recent literature review on sustainability from a general perspective in Software Engineering can be found in~\cite{VENTERS2023107316}.

Beyond all these works that face the challenge of conceptually characterizing the sustainability and/or integrating it into the lifecycle of BPM, our work pays special attention to those approaches that provide specific metrics and frameworks to measure sustainability within BPs, and/or propose prescriptive models that redesign BPs in order to improve them from a sustainability point of view. The following subsections analyze these related works in detail.  

\subsection{Measuring sustainability in BPM}

Process measurement needs to accommodate sustainability-related factors to enable the supervision of achieving sustainability goals and it fosters transparency and awareness crucial for engaging employees across the organization. Hence, it is imperative to develop the necessary measurement techniques and establish IT infrastructures for data collection and comprehensive monitoring of sustainability initiatives.
Measuring sustainability in BPM involves the development and application of various metrics and frameworks. The authors in~\cite{Kohlborn2014} proposed a set of sustainability indicators specifically tailored for BPM, addressing the triple bottom line of sustainability. Other works, e.g.,~\cite{Loos2017,DBLP:conf/bpm/CouckuytLB17,cleven2012managing,zeise2012measurement}, have focused on developing methodologies for assessing the sustainability impact of BPs through life cycle analysis and environmental performance metrics.

The adoption of sustainability performance measurement systems, such as the Global Reporting Initiative (GRI) and the Sustainability Balanced Scorecard (SBSC), has been explored extensively in the BPM context~\citep{Schaltegger2006}. For instance, the authors highlighted the applicability of SBSC in aligning BPs with sustainability goals, providing a structured approach to evaluate sustainability performance at different process levels~\citep{schalteger2006mapping}.
Other works provide more specific solutions to measure values of one sustainable dimension, especially the environmental one. Ghose et al.~propose to annotate, at design time, each task within a BPMN model with the amount of carbon emissions it emits when being executed~\citep{DBLP:journals/ajis/GhoseHHL09}. In addition, these annotations are accumulated step by step over a sequentially ordered pair of tasks to know, at a given point in a process design, the set of contingent measures, each corresponding to an alternative path from the Start Event to that point. This work also proposes the interesting notion of using a resource atomically or shared in the execution of a BP. Houy et al.~extend Event-driven Process Chains (EPCs) to annotate process activities with ratios for, e.g., consumption of energy or fuel and production of carbon dioxide, which are not to be exceeded during process execution, and which can be monitored~\citep{houy2012advancing}. Process modeling needs to incorporate sustainability-related concepts of business activities, such as carbon emissions, energy consumption, and paper consumption. By integrating these sustainability metrics, businesses can better understand and optimize their processes to reduce environmental impact.

As we can see, none of the above-analyzed works face the measurement of BP sustainability in a holistic and systematic manner, as we aim to do in our approach by considering the impact that an existing BP has on multiple sustainability dimensions, and defining indicators that can be used to create metrics that measure sustainability effects, with or without the involvement of IoT. 

\subsection{Improving sustainability in the BPM} \label{sec:sustainability-in-bpm}
Improving sustainability in BPM involves redesigning processes to reduce negative environmental impacts, enhance social benefits, and maintain economic viability. As discussed in~\cite{Seidel2017}, research efforts should focus attention on impactful solutions that focus on translating conceptualization and analysis into design and impact. This includes strategies such as resource-efficient process design, waste minimization, energy consumption reduction, etc.

Several works have already taken some steps in that direction. For instance, Ardagna et al.~propose a layered formal framework to reduce energy consumption in BPs~\citep{ardagna2008active}; Wesumperuma et al.~present a framework to systematically optimize a BP with multiple-dimensions and consider greenhouse gas emission mitigation as one~\citep{wesumperuma2011framework}; and Mancebo et al.~apply gamification to BPMS systems to improve their sustainability, engaging users to be more environmentally friendly in their daily work~\citep{mancebo2017bpms}. In~\cite{lubbecke2017sustainability}, 26 Ecological Process Patterns in BPM were introduced. These patterns can be used for the improvement of existing processes or for the design of new processes in due consideration of ecological goals such as the reduction of resource consumption during the executing of these processes.

With the purpose of sustainability-oriented process analysis and re-design, Klessascheck et al.~introduce a framework that extends the BPM life cycle by the use of Life Cycle Assessment (LCA) for sustainability analysis in combination with Activity based Costing (ABC)~\citep{klessascheck2025sopa}. In the same way, Larsch et al.~propose a process model that integrates the sustainability model into the BPM lifecycle for improvement of sustainability~\citep{larsch2017integrating}. The process consists of seven steps executed in sequence with a feedback loop at the end for continuous process improvement. This proposed process provides a solid basis for identifying, discussing and implementing sustainability-related improvements of typical BPs, either based on a specifically chosen sustainability dimension or on a specific BP. However, it does not provide a conceptual integration of BPM and sustainability and it does not consider IoT devices as potential process performers or for analyzing sustainability. Moreover, the catalog developed by the authors to be considered in the first step (\emph{Brainstorming}) only covers two exemplary BPs and some rather generic advice suggesting exemplary activities to improve a chosen dimension. The sustainability analysis of existing BPs and their activities is not part of the proposed process, which limits its capabilities of identifying opportunities for sustainability improvements in a more systematic way than just a brainstorming session. We aim to address this aspect by providing a conceptually well-founded integration of BPM with sustainability and IoT in a methodology to systematically analyze and improve BPs.

\section{Conceptual Model} \label{sec:framework}

\begin{figure}
    \centering
    \includegraphics[width=0.95\linewidth]{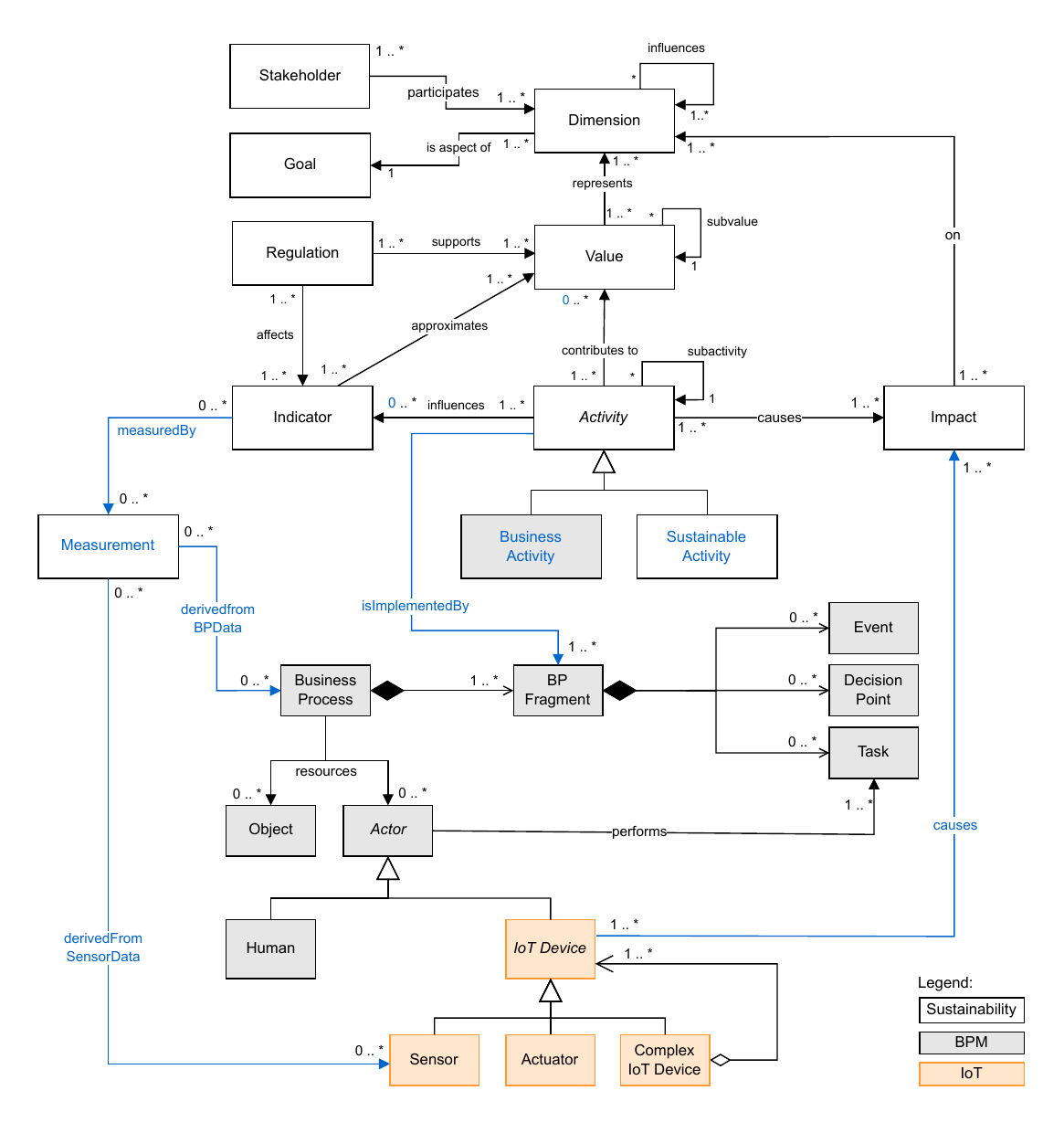}
    \caption{Sustainability-aware, IoT-enhanced BP metamodel (novel, bridging concepts highlighted in blue)}
    \label{fig:metamodelo-extended}
\end{figure}

This section introduces a conceptual model, depicted in~Fig.~\ref{fig:metamodelo-extended}, defining the essential concepts and their relationships to address sustainability within IoT-enhanced BPs (hereafter \emph{Sustainability-aware, IoT-enhanced BP metamodel}). The model enables entities and companies to incorporate sustainability into the design of their IoT-enhanced BPs, evaluate sustainability outcomes, and identify opportunities for process improvement to enhance overall sustainability. 

The Sustainability-aware, IoT-enhanced BP metamodel is built on the sustainability model presented in~\cite{DBLP:conf/aosd/PenzenstadlerF13} (cf.~Section~\ref{subsec:sustainability}). Although this sustainability model is well-suited to the context of software engineering, it requires the incorporation of additional concepts to effectively capture and specify sustainability requirements within the domains of BPM and IoT that frame this work (cf.~Section~\ref{subsec:bpmandiot}).
Therefore, the presented metamodel aims to establish the key concepts of sustainability in IoT-enhanced BPs and the rules that govern them by integrating the fields of sustainability (white boxes), BPM (gray boxes) and IoT (orange boxes), and introducing new concepts to bridge them (blue-colored elements) respectively in Fig.~\ref{fig:metamodelo-extended}. We will illustrate each concept using an excerpt of the running example from the tourism domain introduced in Section~\ref{sub:runningexample}.

\subsection{Sustainability}
\label{subsec:sustainability}
The sustainability aspects of our proposed metamodel (represented by white boxes in Fig.~\ref{fig:metamodelo-extended}) are based on the generic sustainability model introduced in~\cite{DBLP:conf/aosd/PenzenstadlerF13}. The model supports the specification of core sustainability concepts and encompasses the following key elements:

\begin{itemize}
    \item \textbf{Goal}. A Goal represents the sustainability objective, making it possible to explicitly integrate sustainability into requirements engineering. Regarding the hotel stay process example, a hotel’s goal might be to minimize the environmental impact of the guest stay. This means designing the business process for managing the stay in a way that it uses fewer resources, reduces waste, and supports sustainable practices. 
    \item \textbf{Dimension}. A Dimension is an aspect or perspective of a goal that relates to sustainability. Five key dimensions are often considered: 

    \begin{enumerate}
        \item \textbf{Individual (Human)}: Personal freedom, dignity, and the ability to act (e.g., offering guests different options to check in)
        \item \textbf{Social}: Relationships between people and groups (e.g., facilitating communication and interactions of guests with hotel personnel according to their preferences)
        \item \textbf{Economic}: Capital growth, investments, and financial operations (e.g., financial costs related to the stay of a hotel guest)
        \item \textbf{Technical}: Maintenance and evolution of systems (e.g., supporting the management of the hotel stay process through I(o)T systems)
        \item \textbf{Environmental}: Use and management of natural resources (e.g., waste and energy consumption during the hotel stay)
    \end{enumerate}
    \item \textbf{Value}. A Value is a moral or natural good that expresses a particular dimension and may span across several dimensions. Regarding the example, \emph{resource conservation} is a value, which has a direct impact on the environmental dimension, but also on the individual dimension, as it may provide convenience for guests (e.g., reducing paper via a digital check-in).
    \item \textbf{Indicator}. An Indicator is a qualitative or quantitative metric used to measure the fulfillment degree of a value. In the example of a paperless check-in/check-out process, a quantitative indicator could be \emph{number of printed pages saved per guest per month} and a qualitative indicator \emph{Guest satisfaction with the digital check-in/check-out experience}. With every indicator, we associate a \emph{Measurement}, which provides a concrete value to derive or calculate an indicator.
    \item \textbf{Regulation}. A Regulation is an optional element that supports or enforces a value. In the example, local regulations may require hotels to handle guest data securely when offering digital check-in and check-out services, relating to privacy and freedom values from the individual dimension, but also encouraging hotels to limit paper use, relating to the resource conservation value from the environmental dimension.
    \item \textbf{Activity}. An Activity in its original meaning is a sustainability action that contributes to a specific value. Its impact is measured by related indicators. We newly distinguish between two types of activities (highlighted in blue), namely \emph{Business Activities} which belong to an organization's core business and existing BPs to generate business value and potentially influence sustainability; and \emph{Sustainable Activities}, which refer to a measure taken to improve a value. These sustainable activities are meant to improve and advance the sustainability of the business activities. In the running example, these would comprise all activities to move from a manual, paper-based check-in/check-out procedure to a completely digitized one. We introduce this distinction between activity types as we aim to also analyze the sustainability-related effects of existing business activities, which may or may not influence one or many indicators and contribute to one or many values.
\end{itemize}

Besides these concepts, the metamodel includes two more sustainability-related concepts adapted from~\cite{larsch2017integrating}, namely \emph{Impact} and \emph{Stakeholder}. The \emph{Impact} represents the potential effects caused directly or indirectly by an activity, which may not necessarily align with the expected outcome. For example, an initiative aimed at reducing energy consumption by automatically regulating the HVAC during the hotel stay could inadvertently and indirectly lead to increased resource consumption elsewhere (e.g.,~caused by the additional IT infrastructure). The \emph{Stakeholder} represents the various individuals or groups involved in a BP who may play a role in addressing sustainability concerns in different ways. For example, the laundry service provider for the hotel can be a key stakeholder, as its practices regarding sourcing materials and resource usage directly impact the sustainability of the BP.

\subsection{BPM and IoT}
\label{subsec:bpmandiot}
The main link between the BPM-related concepts (gray boxes) and sustainability is via the concept of an \emph{Activity}, which is either part of the existing BP or a sustainable activity. As discussed above, an activity may influence sustainability-related indicators and may contribute--positively or negatively--to different values related to the sustainability dimensions~\citep{albert2024sustainability}. As we focus on the linking of sustainability with the BPM domain, we newly assume that an activity is \emph{implemented by} (highlighted in blue) one or more BP fragments of a BP. Typically, these \emph{BP fragments} comprise a sequence of events, decision points, and tasks~\citep{DBLP:books/sp/DumasRMR18}. Tasks are performed by actors, traditionally in the form of humans or--to provide the novel link with the IoT domain (boxes in orange)--supported by \emph{IoT Devices}, which are either sensors, actuators or complex compounds of sensors and actuators sensing and influencing the physical world~\citep{bauer2013iot} (e.g,~a smart HVAC system that automatically regulates thermal comfort based on environmental sensor readings and presence of the hotel guest).

We introduce a second novel link in this metamodel to bridge the sustainability and IoT fields (boxes in orange), namely the capability of sensors to be used for deriving measurements of sustainability indicators. These indicators can be derived from different data sources, either from data collected from inside the organization (e.g., data from the own BPM systems or from IoT sensors~\citep{bauer2013iot}) or from external sources (e.g., weather prediction). 
For example, the indicator \emph{energy usage} is based on a measurement of \emph{x kilowatt-hours (kWhs)}. A measurement is derived from one or multiple sensors (e.g., an energy meter) and/or BP data collected by an information system (e.g., the number of hotel guest check-ins per hour to derive throughput and efficiency of the check-in). Note that the use of IoT devices, either as active actors to execute tasks in a BP, or as sensors measuring sustainability indicators, will also cause an impact on sustainability (e.g., newly deployed sensors consume energy) and therefore should also be considered~\citep{albert2024sustainability}. 

\section{Methodology} \label{sec:methodology}

\begin{figure}
    \centering
    \includegraphics[width=1\linewidth]{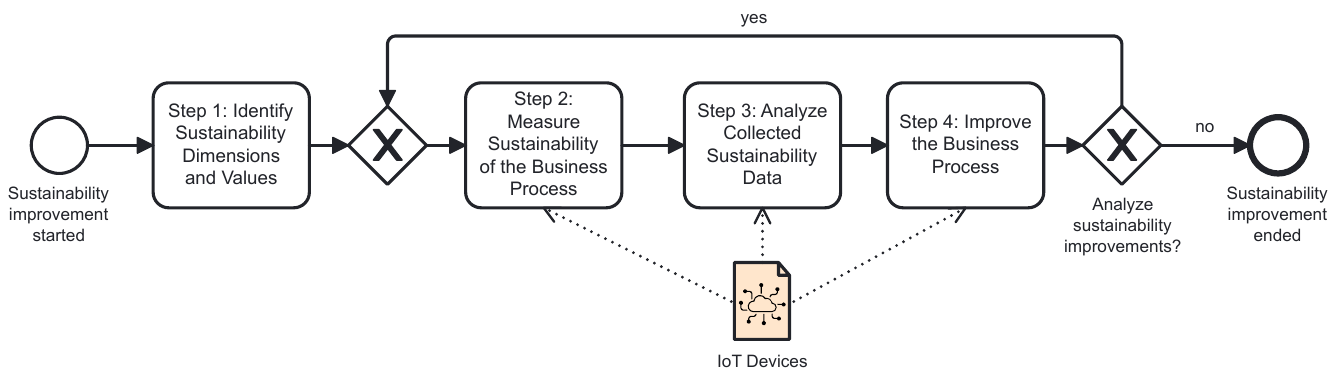}
    \caption{The core steps of the methodology with highlighted IoT involvement}
    \label{fig:Metodologia}
\end{figure}

This section proposes a methodology--drawing on the work described in~\cite{larsch2017integrating}--that provides a structured foundation to support organizations in achieving sustainability goals by the support of IoT devices. Building on the concepts and relationships defined in the conceptual model (cf.~Section~\ref{sec:framework}), the methodology helps organizations to analyze current BPs according to one or more sustainability dimensions (cf.~RQ--1 in Section~\ref{sec:problemStatement}) and to improve these models (cf.~RQ--2 in Section~\ref{sec:problemStatement}) by incorporating new actions to enhance BP performance across such dimensions. Figure~\ref{fig:Metodologia} depicts the four core steps of the proposed methodology. A key feature of the methodology is the integration of IoT technologies in some of its steps to support dynamic, evidence-based sustainability enhancements. In addition, to ensure a comprehensive and effective review of BPs in terms of sustainability, it is essential to involve a diverse set of organizational roles in this process. We advocate for a  multidisciplinary team capable of ensuring that the analysis and improvement of BPs are robust, feasible, and aligned with the organization’s sustainability commitments. In our proposal we envision a multidisciplinary team including the following roles:

\begin{itemize}
        \item \textit{Senior executives and strategic leaders (Senior-executive)} who can provide the vision and strategic alignment necessary to embed sustainability goals into core business objectives.
        \item \textit{Process owners and functional managers (Process-owner)} who can contribute by detailing operational knowledge and identifying practical opportunities for improvement. 
        \item \textit{Sustainability experts (Sustainability-expert)} who can assess the impacts of organizational activities in terms of sustainability across their various domains, establish meaningful indicators, and ensure alignment with standards and regulations.  
        \item \textit{IoT experts (IoT-expert)} who design, deploy, and optimize the overall IoT infrastructure.
        \item \textit{Data engineers (Data-engineer)} who manage the efficient acquisition, processing, analysis, and storage of data.
        \item \textit{Frontline employees and supervisors (employees)} who can offer valuable insights into day-to-day practices and operational realities. 
        \item \textit{External stakeholders (external-stakeholder)} who play a critical role in sustainability when processes extend along the supply chain (e.g., key suppliers, auditors, or sustainability consultants).
        \item \textit{Data collectors (Data-collector)} who are responsible for gathering data for analysis purposes. This data can be acquired either automatically or manually and may include: 1)~sensors/IoT devices to support subsequent measurement, 2)~data from the information systems related to the BP executions (e.g., BPM systems), and 3)~BP-related data collected from frontline employees, external stakeholders (e.g., via surveys).
\end{itemize}
 
Sections~\ref{sub:step1} to~\ref{sub:step4} delve into the four steps of the methodology, developing for each task a detailed description of what it entails, who the main actor responsible for its execution is (note that any of the roles introduced above can also be involved), what the input and output artifacts are, and a reference to the running example presented in Section~\ref{sub:runningexample}. 
  
\subsection{Step 1: Identify sustainability dimensions and values} \label{sub:step1}
\begin{figure}
    \centering
    \includegraphics[width=1.0\linewidth]{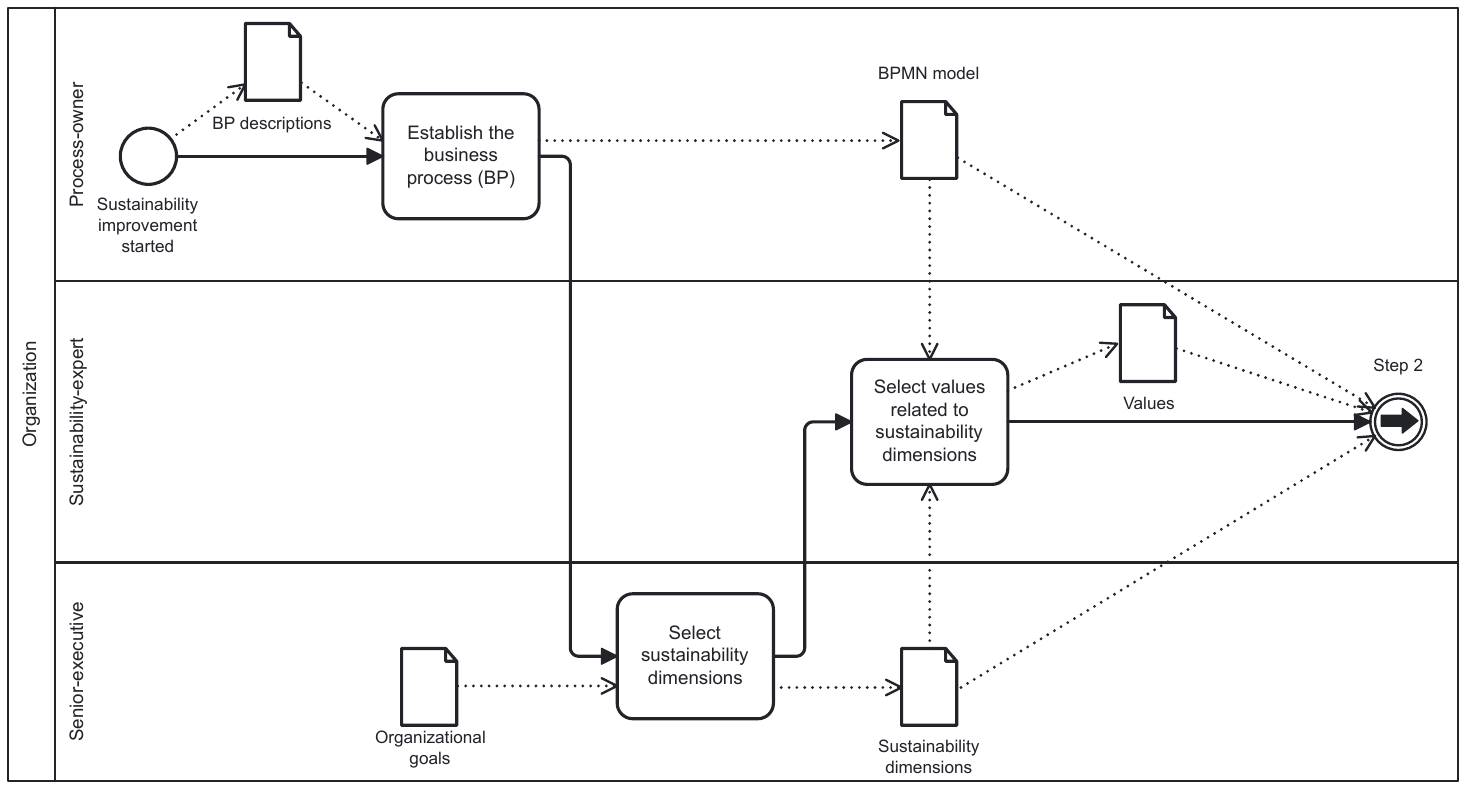}
    \caption{Actors, tasks and artifacts of Step 1 in the methodology}
    \label{fig:Step1Metodologia}
\end{figure}

The initial step of the proposed methodology focuses on the identification and selection of the sustainability \textit{dimensions} and corresponding \textit{values} that are pertinent to the evaluation and improvement of a given BP. This step serves as the foundation for aligning BP improvements with sustainable development \textit{goals}. Figure~\ref{fig:Step1Metodologia} illustrates the tasks associated with this phase.

\begin{enumerate}

  \item Task: Establish the business process (BP).
     \begin{itemize}
        \item Description: The methodology starts by establishing the \emph{business process} (BP) to be analyzed based on descriptions of the BP. It is important to note that, depending on who has represented the BP and for what purpose, the BP may differ both in the level of detail in which it is described and in its form (whether expressed in unstructured natural language or in a more structured manner, e.g., as a BPMN model). In any case, it is essential to ensure that the stakeholders have a clear and comprehensive understanding of the process under analysis. As output of this step, we assume that there is a BPMN model available or created of the BP to be analyzed for sustainability improvements.
        \item Main Actor: Process-owner.
        \item Input Artifacts: BP description(s).
        \item Output Artifacts: Established BP as BPMN 2.0 model.
        \item Running example: Hotel stay process description and model (cf.~Section~\ref{sub:runningexample}) provided by the hotel manager.
    \end{itemize}
    
  \item Task: Select Sustainability Dimensions.
    \begin{itemize}
        \item Description: Once the \textit{BP} is clearly established, the next step involves selecting the sustainability \textit{dimension(s)} that are most relevant to the organization’s strategic priorities. Typically, these span across the five sustainability dimensions (cf.~Section~\ref{subsec:sustainability}) and the selection is performed by the senior executive(s) based on the organizational \textit{goals} which may be driven by regulatory demands, stakeholder interests or expectations, and strategic objectives.
        \item Main Actor: Senior-executive.
        \item Input Artifacts: Organizational goals.
        \item Output Artifacts: A list of selected sustainability dimensions relevant for process measurement and/or evaluation.
        \item Running example: The general manager (senior-executive) has set organizational goals focused on \emph{reducing waste} and \emph{improving the guest experience}. Based on these goals, he suggests to address the \emph{environmental dimension} related to waste management, and the \emph{individual dimension} associated with user perception and well-being.
    \end{itemize}

\item Task: Select Values Related to Sustainability Dimensions.
    \begin{itemize}
        \item Description: In this task, the BPMN model and the selected sustainability dimensions are processed to identify specific \emph{values} that are measurable and susceptible to improvement. To this end, the BP has to be analyzed with the objective of generating a preliminary list of values associated with the chosen dimensions, which serves as a foundation for further refinement. In this step, the sustainability expert also assesses the selected values for potential conflicts and contradictions in achieving the sustainability goals. This requires a careful trade-off analysis and prioritization of sustainability values to be improved as well as an impact analysis of dependent sustainability values~\citep{ruhe2003trade}.
        \item Main Actor: Sustainability-expert.
        \item Input Artifacts: BP model and the list of selected sustainability dimensions.
        \item Output Artifacts: A list of values relevant for measurement and improvement.
        \item Running example: Given the BP model, the selected dimensions and established goals, the hotel's sustainability expert suggests that \emph{resource efficiency} and \emph{guest satisfaction} should be prioritized as values for the environmental and individual dimensions, respectively.
    \end{itemize}

\end{enumerate}  

\subsection{Step 2: Measure sustainability of the Business Process}  \label{sub:step2}
\begin{figure}
    \centering
    \includegraphics[width=1\linewidth]{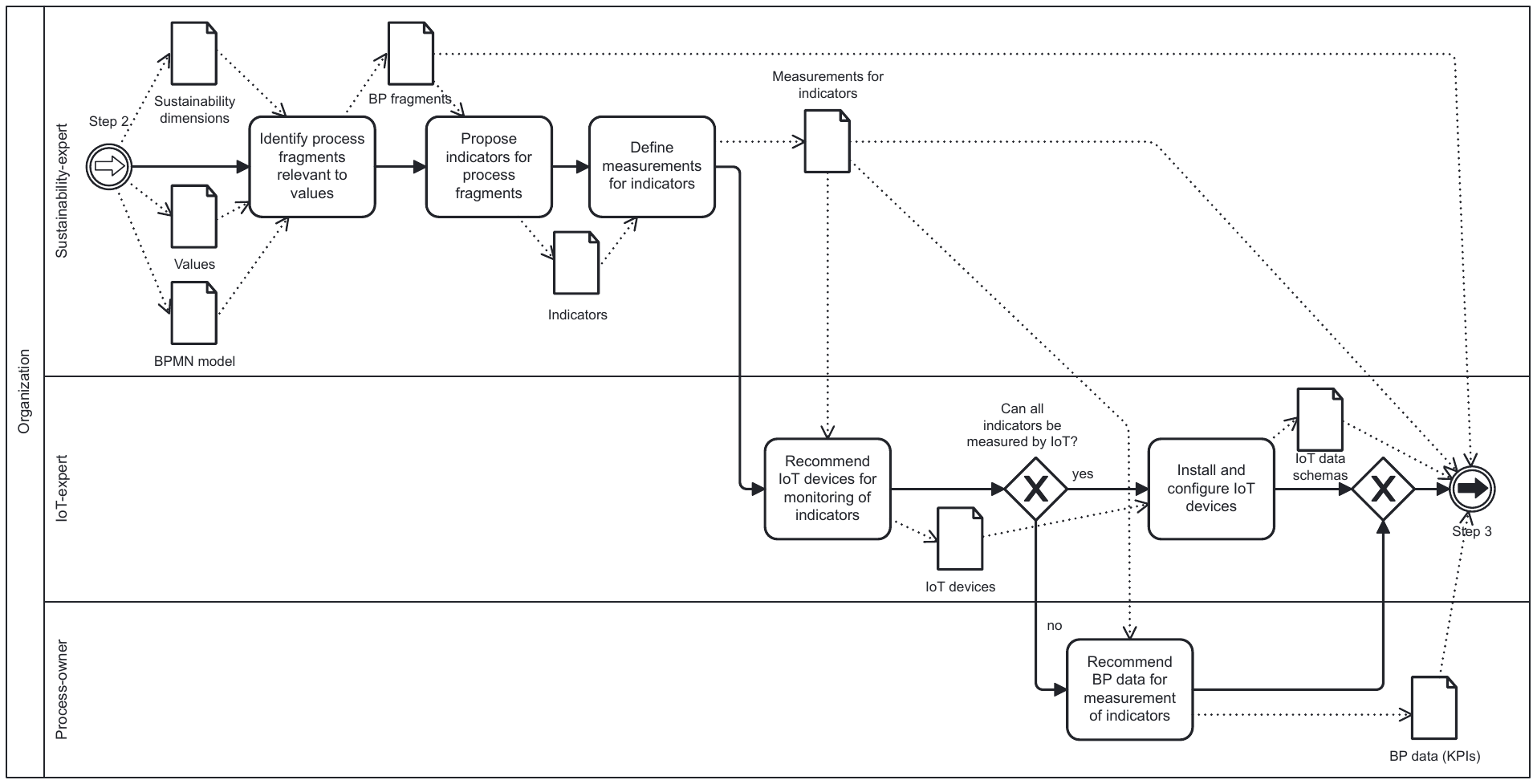}
    \caption{Actors, tasks and artifacts of Step 2 in the methodology}
    \label{fig:Step2Metodologia}
\end{figure}

The second step of the methodology focuses on measuring the provided BP in accordance with the sustainability values identified in Step 1. This step involves isolating relevant \emph{BP fragments}, defining \emph{indicators}, and implementing mechanisms for data acquisition. Figure~\ref{fig:Step2Metodologia} outlines the sequence of tasks.

\begin{enumerate}
    \item Task: Identify Process Fragments Relevant to Selected Values.
    \begin{itemize}
        \item Description: This task entails identifying segments within the BPMN model (i.e., \emph{BP fragments}) that are likely to influence the sustainability values identified in step 1, either positively or negatively. It is important to note that multiple fragments may contribute simultaneously to the same value. Therefore, their associated indicators must be jointly considered by the sustainability expert, taking into account possible dependencies to avoid double counting when evaluating the overall value.
        
        \item Main Actor: Sustainability-expert.
        \item Input Artifacts: The BP model, the selected dimensions, and values.
        \item Output Artifacts: A set of \textit{BP fragments} relevant to sustainability values.
        \item Running example: Two fragments are identified by the hotel's sustainability expert (cf.~Fig.~\ref{fig:task2.1-fragments}): one containing the ``Verify reservation'', ``Request documents'', and ``Select available room'' tasks for the \emph{guest satisfaction} value (Fragment 1); and a second one containing the ``Hand over physical key'' and ``Switch on/off appliances and HVAC'' tasks for the \emph{resource efficiency} value (Fragment 2).
    \end{itemize}

    \begin{figure}
    \centering
    \includegraphics[width=1.0\linewidth]{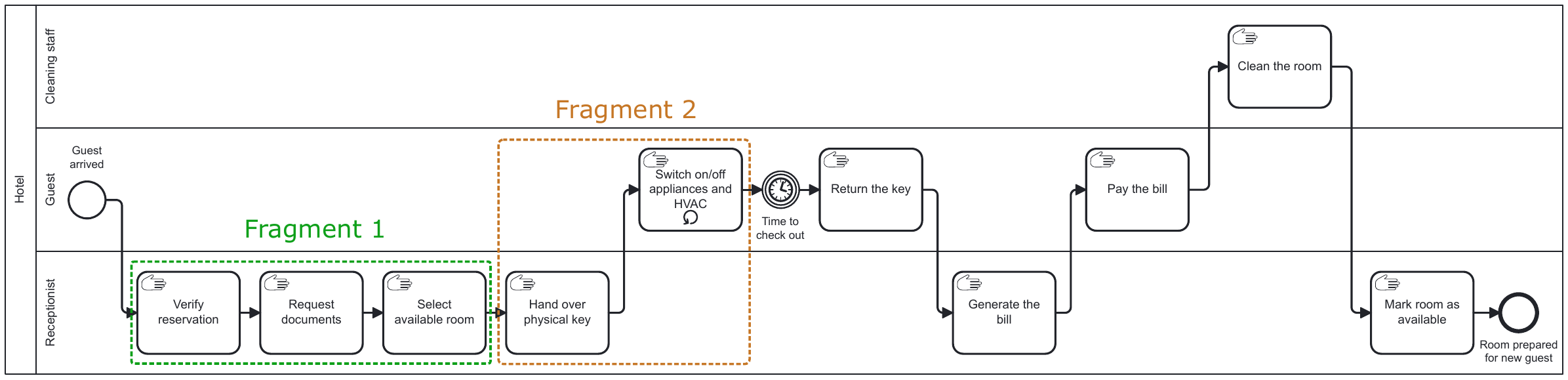}
    \caption{Sustainability-related fragments identified in the Hotel Stay example process}
    \label{fig:task2.1-fragments}
    \end{figure}
 
    \item Task: Propose Indicators for Process Fragments.
    \begin{itemize}
        \item Description: For each identified process fragment, a corresponding set of sustainability \emph{indicators} should be proposed by the sustainability expert. These indicators quantify the BP fragment's impact on the selected sustainability values and allow structured measurement. These indicators can be self-defined by the organization or based on standard indicators for the specific domain/problem (e.g., following an ISO standard or regulation). The definition of each indicator is accompanied by a custom set of value ranges that classify the performance level (e.g., excellent, acceptable, moderate, or poor), providing a framework for interpreting and comparing the associated metrics. The process owner evaluates the proposed list of indicators and selects those that can be realistically implemented in the process.
        \item Main Actor: Sustainability-expert.
        \item Input Artifacts: A set of BP fragments.
        \item Output Artifacts: A set of indicators for each BP fragment.
        \item Running example:  For the first fragment the self-defined \emph{Manual Check-in Fluency Index (MCFI)} is proposed, which reflects how smooth and pleasant the check-in procedure with the receptionist is from the guest's perspective and it focuses on: the perception of time (not just its actual duration), the clarity and simplicity of the procedure, and the absence of discomfort or friction. Custom value ranges for this indicator could be: 0.76-1 (excellent), 0.6-0.75 (acceptable), 0.26–0.5 (moderate: requires review), and $<= 0.25$ (poor: requires immediate action). For the second fragment, the self-defined \emph{Carbon Footprint Index per Day (CFID)} is proposed, which is meant to measure the environmental impact (in kg of CO$_2$e) per day associated with the stay of one hotel guest, including the handover of a physical space (e.g., a hotel room), and the energy and material resource usage during the stay. Custom value ranges for this indicator could be: 0–2.2 (excellent), 2.21–3.5 (acceptable), 3.51–6 (moderate: requires review), and $> 6$ (poor: requires immediate action).        
    \end{itemize}   

    \item Task: Define Measurements for Indicators.
    \begin{itemize}
        \item Description: This task specifies how an indicator can be obtained, i.e., it specifies the measurement and/or formula used to calculate its value, along with the data required for the computation. As discussed above, these formulas can be referring to self-defined or standard measurements. The formulas will be later populated with data obtained from different available sources (e.g., from the BP, IoT devices, etc.).
        \item Main Actor: Sustainability-expert.
        \item Input Artifacts: A list of indicators.
        \item Output Artifacts: A measurement to compute each indicator. 
        \item Running example: To measure the \emph{MCFI indicator}, the following self-defined formula is proposed, producing values between 0 and 1, where higher values indicate a more satisfactory check-in experience. The MFCI is calculated based on the mean of the individual components over a given observation period: 

\[
\text{MCFI} = \frac{\overline{S'} + (1 - \overline{F'}) + \overline{P'}}{3}
\]

\begin{itemize}
            \item 
            \begin{description}
                \item $\overline{S'}$ represents the mean of the normalized guest satisfaction score ($S'=S/10$) values for all guests obtained from the surveys over the specified observation period. The guest satisfaction score $S$ ranges from~0 to~10 and is interpreted as follows: 0–2.5~indicates a poor user satisfaction, 2.6–5~a fair user satisfaction, 5.1–7.5~a good user satisfaction, and~7.6–10 an excellent user satisfaction.
            \end{description}
            \item 
            \begin{description}
                \item $\overline{F'}$ represents the mean of the normalized number of frictions reported ($F'=F/F_{max}$) for all guests, where $F$ is the number of frictions and $F_{max}$ is the maximum number of frictions reported by a guest in the observation period.
            \end{description}
            \item 
            \begin{description}
                \item $\overline{P'}$ represents the mean of the normalized time values as perceived by a guest ($P'=P/10$) obtained from the surveys for all guests over the specified observation period. The ratings of perceived time $P$ range from~0 to~10 and are interpreted as follows: 0–2.5~indicates a poor perception, 2.6–5~a fair perception, 5.1–7.5~a good perception, and 7.6–10~an excellent perception.
            \end{description}
        \end{itemize}

To measure the \emph{CFID indicator} representing the carbon footprint of one hotel guest per day, the following formula is proposed (per hotel room and stay): \[\text{CFID} = \frac{(E_{\text{appliances}} \times EF_{\text{energy}}) + (E_{\text{HVAC}} \times EF_{\text{energy}}) + EM_{\text{material}}}{(N \times D)}
\] where:
        \begin{itemize}
            \item 
            \begin{description}
                \item[$E_{\text{appliances}}$:] Energy consumed when actively using appliances in a room, e.g., lights or TV (kWh).
            \end{description}
            \item 
            \begin{description}
                \item[$E_{\text{HVAC}}$:] Energy consumed by HVAC systems (kWh) in the room.
            \end{description}
            \item 
            \begin{description}
                \item[$EF_{\text{energy}}$:] Environmental emission factor (kg CO\textsubscript{2}e/kWh), depending on the energy source used.
            \end{description}
            \item 
            \begin{description}
                \item[$EM_{\text{material}}$:] Emissions (in kg CO$_2$e) associated with the manufacturing or replacement of the physical key cards (if a new one is needed). This factor is provided by the manufacturer of the key cards.
            \end{description}
            \item
            \begin{description}
                \item[$N$:] Number of guests per room.
            \end{description}
                        \item
            \begin{description}
                \item[$D$:] Number of days of the stay.
            \end{description}
        \end{itemize}
    \end{itemize}
Note that the calculation and analysis of the CFID indicator in later steps will also be referring to its average value for all guests over a given period of time.
    
    \item Task: Recommend IoT Devices for Monitoring of Indicators.
    \begin{itemize}
        \item Description: Based on the proposed measurements, this task identifies suitable \emph{IoT devices} capable of capturing the data required for achieving the measurement calculations that will provide a value for the respective indicators. Note that in this step, it is within the responsibility of the IoT expert together with the sustainability expert to also assess the impact of installing and operating the recommended IoT devices. IoT devices generate initial and operational costs that might negatively impact relevant and associated sustainability dimensions~\citep{albert2024sustainability}.
        \item Main Actor: IoT-expert.
        \item Input Artifacts: A measurement to compute each indicator.
        \item Output Artifacts: A recommended list of IoT devices used to monitor indicators.
        \item Running example: To monitor the MCFI indicator, it is proposed to collect data via questionnaires that guests are asked to complete after finishing the check-in procedure. Regarding the CFID indicator, the following IoT devices are recommended to monitor it:
        \begin{itemize}
            \item Shelly Plug S for appliances: Smart plugs with energy monitoring for measuring electricity consumption in sockets and devices. 
            \item Sensibo Air Pro for HVAC: Smart controllers for HVAC systems to monitor and control energy usage in air conditioning and heating.  
        \end{itemize}
        
    \end{itemize}

    \item Task: Install and Configure IoT Devices.
    \begin{itemize}
        \item Description: In this task, the recommended IoT devices are physically deployed within the business environment to enable monitoring. The configuration process involves setting data collection parameters, establishing connectivity settings, and integrating the devices with monitoring platforms and databases.
        \item Main Actor: IoT-expert.
        \item Input Artifacts: IoT devices proposed for monitoring the different indicators related to the process fragments.
        \item Output Artifacts: Data schema representing the structure and format of the data from installed IoT devices. 
        \item Running example: In addition to installing the previously recommended IoT devices to calculate the CFID indicator, the data schema from these should also be clearly defined. For example, from the Shelly Plug S device we would get data containing: \texttt{device\_name} (assigned name for the plug), \texttt{timestamp} (date and time of the data capture), \texttt{instantaneous\_power\_w} (current power consumption), \texttt{device\_temperature\_c} (internal temperature of the plug), \texttt{device\_state} (indicates whether the device was ON or OFF at that time), and \texttt{created\_at} (when the reading was logged into a database). It is recommended to integrate the proposed IoT devices with Home Assistant as a management and analytics platform to facilitate the calculation of the CFID.
        \end{itemize}

    \item Task: Recommend BP data for Measurement of Indicators.
    \begin{itemize}
        \item Description: In case not all indicators can be effectively measured by IoT devices, as determined in Task~2.4, the process owner and the data engineer must locate alternative data sources to allow the calculation of the proposed indicators. These indicators primarily come from data related to the existing BP (e.g., KPIs), from related information systems (e.g., ERP systems), from manual records (e.g., forms), or from external sources. In any case, these sources must ensure that the collected data is reliable, available, and current.
        \item Main Actor: Process-owner.
        \item Input Artifacts: A list of non-IoT related measurements for indicators.
        \item Output Artifacts: A recommended list of sources for BP-related data to cover the measurements of indicators.
        \item Running example: To calculate the MFCI, guest satisfaction surveys are identified as valid sources of data for these missing BP data. These surveys are available for guests to be filled out right after entering their hotel rooms to collect data related to their experience at the hotel (e.g., fluency and waiting time during the check-in). 
    \end{itemize}  
\end{enumerate}

\subsection{Step 3: Analyze collected sustainability data}  \label{sub:step3}
\begin{figure}
    \centering
    \includegraphics[width=1.00\linewidth]{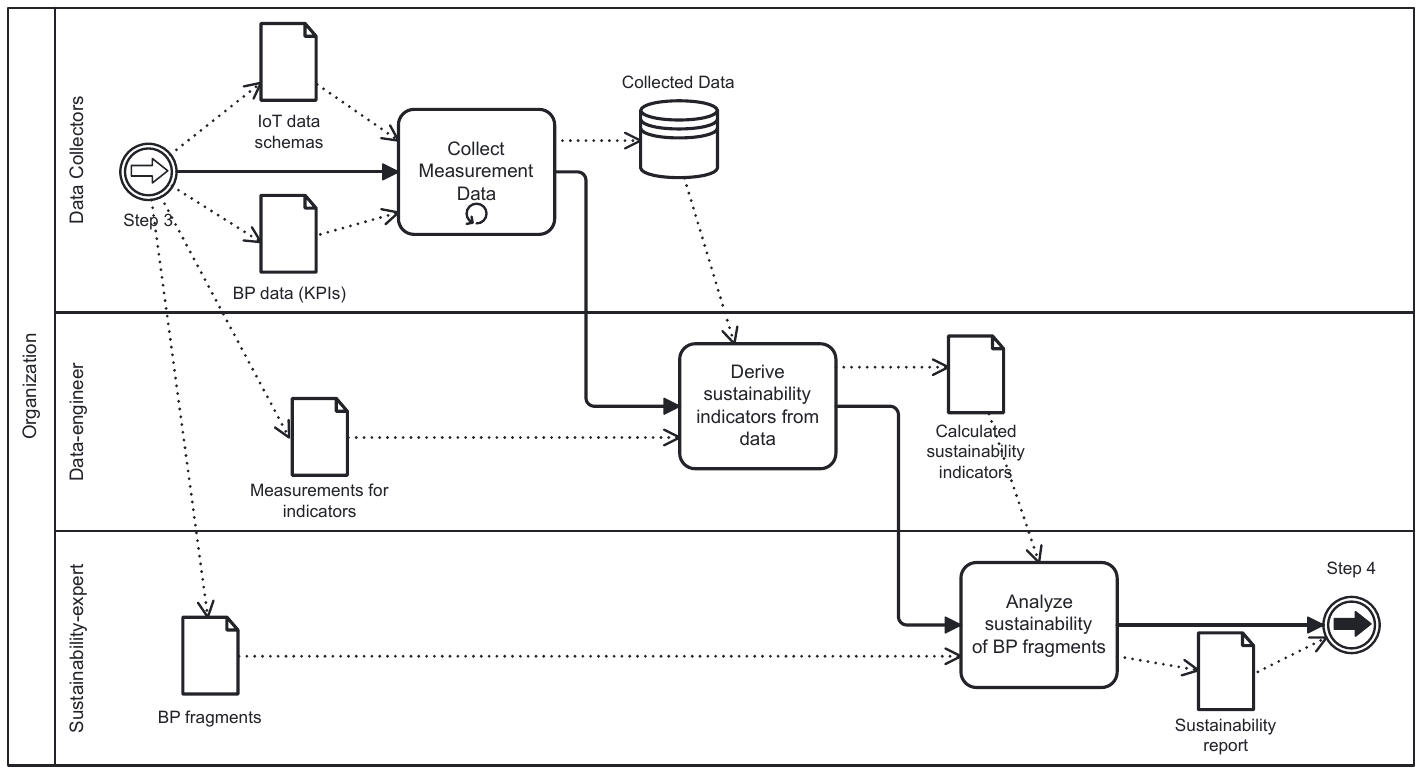}
    \caption{Actors, tasks and artifacts of Step 3 in the methodology}
    \label{fig:Step3Metodologia}
\end{figure}

The third step involves acquiring and preparing the data from the IoT devices and business processes, and using them to calculate the sustainability indicators defined earlier. This step transforms raw data into insights for decision support as shown in Fig.~\ref{fig:Step3Metodologia}.

\begin{enumerate}
    \item Task: Collect Measurement Data.
    \begin{itemize}
        \item Description: This task is responsible for acquiring data via different data collectors during a specified observation period. 
        \item Main Actors: Data-collector.
        \item Input Artifacts: IoT data schemes, BP data, other data, and observation period.
        \item Output Artifacts: Collected data in a database. 
        \item Running example: Data from 122 hotel stays have been recorded over a three-month observation period across three short-stay rooms. During this period, the Home Assistant system collects and stores data generated by the installed IoT devices, such as smart plugs and smart HVAC systems. The filled-in surveys from the hotel guests are manually entered into a database.
    \end{itemize}
    
    \item Task: Derive Sustainability Indicators from Data.
    \begin{itemize}
        \item Description: This task prepares the data collected from sensors and other relevant sources into actionable information to derive the sustainability-related \emph{indicators} as defined in Task~2.3. 
        \item Main Actor: Data-engineer.
        \item Input Artifacts: Collected data and measurements for indicators. 
        \item Output Artifacts: Sustainability indicators calculated from data.
        \item Running example: the MCFI is calculated using the guest survey responses. According to the surveys recorded in the observation period, 94 of the check-ins were completed without reported issues, while the remaining 28 experienced some form of friction--including delays in key delivery (12 cases), incorrect room assignments~(7), and missing guest documentation~(9). The mean, normalized values obtained are: $\overline{S'} = 0.4$ for guest satisfaction; $\overline{F'} = 0.39$ for frictions; and $\overline{P'} =  0.38$ for perceived time; which gives a resulting MFCI score of 0.46 for the three-month observation period. 
        
        Regarding the CFID, before calculating it, the data collected over the three-month period must be properly prepared. In this case, the data engineer must convert the energy readings provided by the Shelly Plug S from Wh to kWh, filter the relevant consumption periods, and calculate the averages per guest and day. The processed, average values are: 2.5 kWh for $E_{\text{appliances}}$; 4.1 kWh for $E_{\text{HVAC}}$; 0.4 kg CO$_2$e/kWh for $EF_{\text{energy}}$; 0.004 kg CO$_2$e for $EM_{\text{material}}$ (since a new key was delivered in 15 out of the 122 cases, assuming $EM_{\text{material}}$ = 0.030 kg CO$_2$e per card); which gives a resulting CFID value of 2.644 kg CO$_2$e per guest per day for the three-month observation period. 
    \end{itemize}

        \item Task: Analyze Sustainability of BP Fragments.
    \begin{itemize}
        \item Description: In this step, the sustainability expert carefully analyzes the calculated sustainability indicators in correlation with the BP fragments to create a comprehensive sustainability report, which is the basis for decision making and process modification in the next step.
        \item Main Actor: Sustainability-expert.
        \item Input Artifacts: Calculated sustainability indicators and BP fragments. 
        \item Output Artifacts: Sustainability report.
        \item Running example: Based on the thresholds defined in Task~2.2, the MCFI score of 0.46 for Fragment~1 indicates a moderately fluent process, but with clear room for improvement. To better understand the root causes, the sustainability expert should analyze the reported frictions and the perceived completion times. 
        
        Although the value obtained for CFID (2.644 kg CO\textsubscript{2}e), associated with Fragment~2, is within the range considered as acceptable, it shows potential for improvement and should be reviewed.
        
    \end{itemize}
    
\end{enumerate}

\subsection{Step 4: Improve the Business Process}  \label{sub:step4}
\begin{figure}
    \centering
    \includegraphics[width=0.95\linewidth]{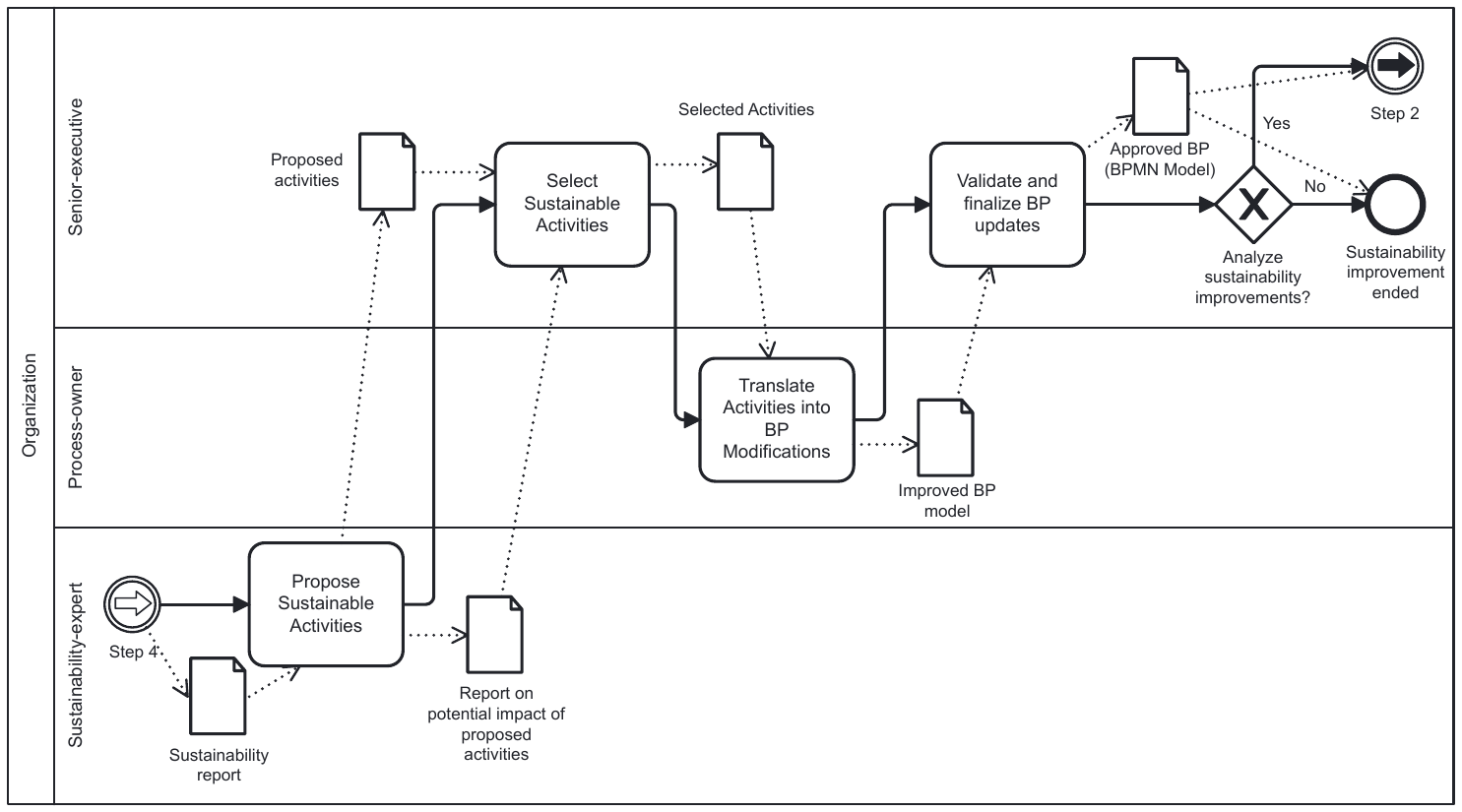}
    \caption{Actors, tasks and artifacts of Step 4 in the methodology}
    \label{fig:Step4Metodologia}
\end{figure}

The final step of the methodology is related to improving the BP based on the analysis performed previously. Figure \ref{fig:Step4Metodologia} details the tasks designed at this stage as well as the roles and the data objects produced and consumed along this step.

\begin{enumerate}
    \item Task: Propose Sustainable Activities.
    \begin{itemize}
        \item Description: In this task, a set of potential \textit{sustainable activities} that could enhance the sustainability performance of the BP are suggested. The activities are derived from the results of prior analyses and identified inefficiencies as provided in the sustainability report from the previous step. This report should also cover the effects of implementing the sustainable activities on other sustainability dimensions derived in earlier analysis (Tasks~1.3 and~2.4).
        \item Main Actor: Sustainability-expert.
        \item Input Artifacts:  Sustainability report. 
        \item Output Artifacts: A list of proposed sustainable activities for improvement.
        \item Running example:  Based on the suboptimal value for the CFID calculated in the previous step, it would be reasonable to propose changes in the BP to \emph{digitize room access control} and \emph{automate the control of the HVAC system}. The same applies to the value obtained for the MCFI indicator, considering a \emph{pre-arrival, digital check-in}, which would include a pre-verification, and simplify document requirements and room selection; and a corresponding \emph{digital check-out and payment}, which would improve the overall obtained results and save additional resources (e.g.,~paper).
    \end{itemize}  

    \item Task: Select Sustainable Activities.
    \begin{itemize}
        \item Description: This task is responsible for reviewing and selecting the most appropriate \textit{sustainable activities} from the previously proposed list. The selection is typically guided by feasibility assessments of the implementation, return on investment, and alignment with organizational goals. The outcome is a subset of \textit{activities} deemed suitable and feasible for implementation in the current process context.
        \item Main Actor: Senior-executive.
        \item Input Artifacts: A list of proposed sustainable activities for improvement.
        \item Output Artifacts: A list of selected sustainable activities for improvement.
        \item Running example: Let us consider that all proposed sustainable activities (cf.~Task~4.1) are selected for the improvement of the hotel stay process.
    \end{itemize} 

    \item Task: Translate Actions into Business Process Modifications.
    \begin{itemize}
        \item Description: This task involves translating the selected sustainability activities into concrete modifications of the BP (model). These may involve: 1)~altering the BP (removing/adding new tasks, events or decision points), 2)~adapting the process logic, 3)~introducing automation (e.g., via IoT devices as actors), or 4)~modifying the infrastructure supporting the process executions and end-users to reflect the intended improvements and ensure alignment with the sustainability objectives.
        \item Main Actor: Process-owner.
        \item Input Artifacts: Selected sustainable activities.
        \item Output Artifacts: Sustainability-improved BP model. 
        \item Running example: In this case, the selected sustainability activities are introduced in the BPMN model as shown in Figure \ref{fig:checkInOutImproved}. We introduce an online check-in for the user, who will receive an automated message 24~hours before check-in. During online check-in using the hotel app, the user provides necessary details and an estimated time of arrival (ETA). After completion, the \emph{smart hotel system} sends a digital key to the user's phone. Shortly before the guest's ETA, the room's HVAC is prepared automatically for guest comfort. The guest unlocks the NFC-enabled door lock with the digital key on the smartphone. During the stay, the \emph{smart room} is able to detect when the guest left the room (e.g., based on data from the door lock, a motion sensor, and time) and regulates HVAC and appliances automatically. On the day of check-out, the guest receives a request on its smartphone to pay the bill and perform the check-out. After the guest completed this process and left the room, the cleaning staff is informed, and the room will be marked as available after cleaning.
    \end{itemize} 

    \begin{figure}
    \centering
    \includegraphics[width=1.0\linewidth]{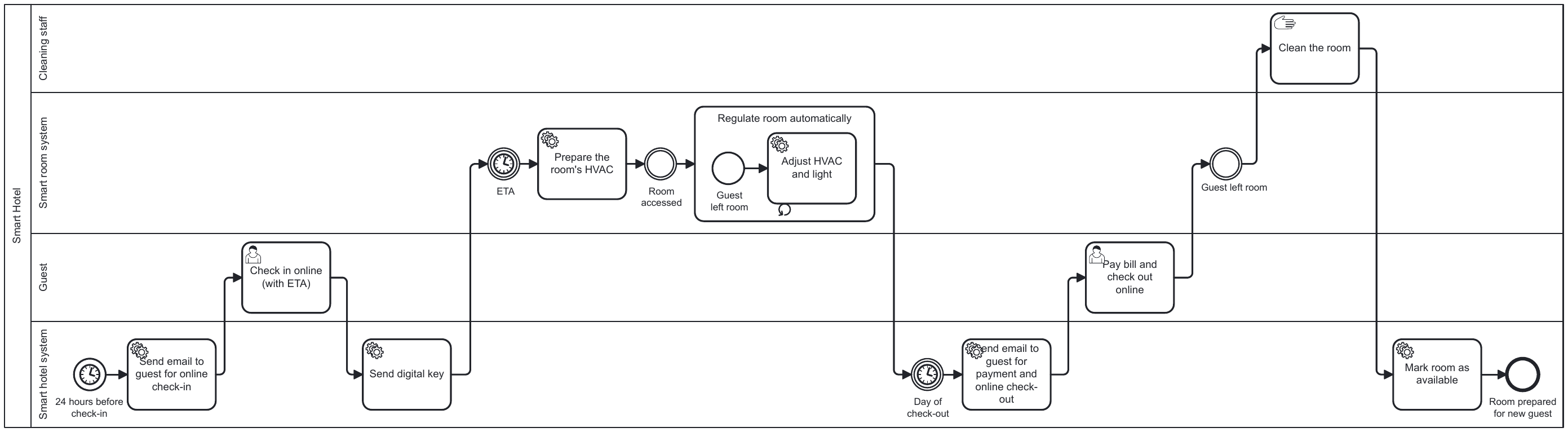}
    \caption{Sustainability-aware, IoT-enhanced BP representing the stay of a guest in a hotel}
    \label{fig:checkInOutImproved}
    \end{figure}

    \item Task: Validate and Finalize Business Process Updates.
    \begin{itemize}
        \item Description: This task focuses on reviewing and refining the proposed modifications to the BP (model) to ensure consistency, feasibility, and alignment with both business and technical constraints. It involves validating correctness and verifying semantic coherence. Finally, the senior executive decides if an analysis of the sustainability improvements (\emph{Impact)} of the modified BP should be conducted. In case the impact should be analyzed, the modified, approved BP will serve as input for Step~2 of the methodology to repeat the measurements and analysis of the relevant sustainability dimensions. If no analysis should be conducted, the process of improving the sustainability of the chosen BP is finished. The result is a sustainability-aware, IoT-enhanced BP. New sustainability improvement initiatives can start anytime by instantiating the proposed methodology as a new process.
        \item Main Actor: Senior-executive.
        \item Input Artifacts: Sustainability-improved BP model.
        \item Output Artifacts: Approved BP (model) with sustainability improvements. 
        \item Running example: The general manager reviews the new version of the BP model and decides to finish the process since it is expected that the applied measures will contribute positively to reach the initially set goals of improving the sustainability of the original BP. Overall, we expect an improved guest experience and reduced waste based on these changes, which are aligned with the sustainability goals. The value \emph{resource efficiency} is achieved through the automated regulation of the HVAC systems and appliances when no guest is present, and through the savings of paper during check-in, check-out and payment, and the savings of plastic associated with the hotel key cards. With the digitized check-in/check-out procedure, we also expect savings in cost for staff. The \emph{guest satisfaction} value should increase through the customized, digital check-in, check-out, payment, and room access via the guest's personal smartphone, which would remove redundant steps (e.g., data entry in forms) and causes of frictions, and reduce waiting times. Note that the introduction of new IT and IoT systems to improve the sustainability of the hotel process regarding certain values may also have negative impact on other dimensions and factors (e.g.,~costs associated with developing and maintaining new software), which need to be analyzed~\citep{albert2024sustainability}. 
    \end{itemize} 
    
\end{enumerate}

\section{Controlled Case Study in Smart Healthcare} \label{sec:case-studies}

To show a proof-of-use of the proposed methodology,
we performed a controlled case study by applying the methodology to a business process based on a real scenario in smart healthcare. This hybrid experimental approach following action research allows us to analyze the feasibility and utility of our approach in a controlled real-world environment, enabling us to evaluate whether the modeling constructs and methodology are expressive and robust enough to describe a representative system with the purpose of analyzing sustainability and introducing IoT enhancements as a proof-of-use~\citep{venable2012comprehensive}. Note that in this case study, the authors of this article applied the suggested methodology to show a proof-of-concept. At several points, different external stakeholders in accordance with the roles introduced in Section~\ref{sec:methodology} were involved. In this case study we collected process execution-related data via identified and newly installed IoT devices, and via questionnaires to assess and improve the sustainability of a process for medical trainees in smart healthcare.

\subsection{Preliminaries}

\begin{figure}
    \centering
    \includegraphics[width=0.75\linewidth]{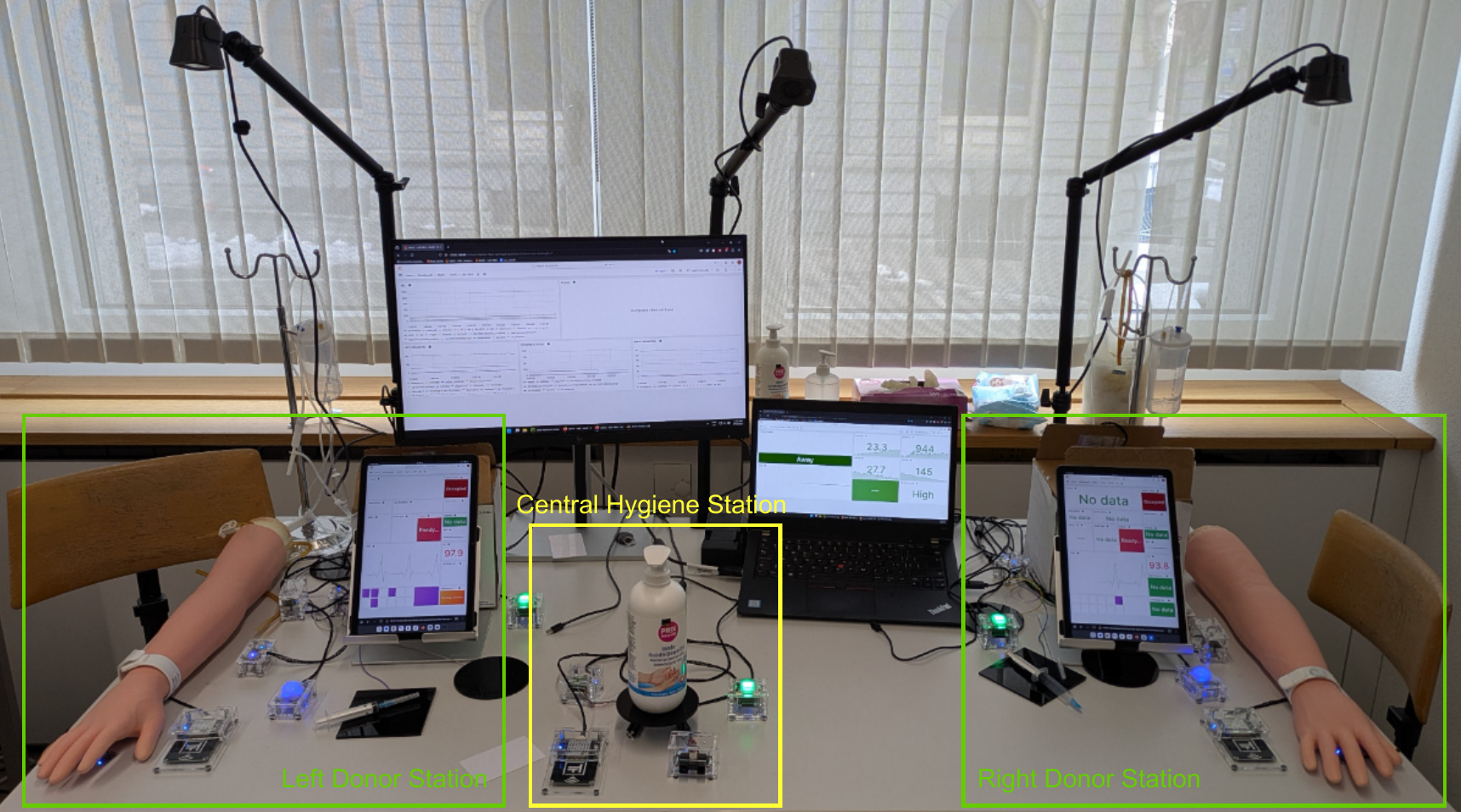}
    \caption{Smart healthcare setup to monitor medical procedures (Task 2.5).}
    \label{fig:iot-setup}
\end{figure}

The case study is set in the domain of smart healthcare. Together with our project partners from the Department of Infection Prevention at the Hospital in St.Gallen (Switzerland), we investigate the use of IoT technologies to monitor and track medical procedures~\citep{franceschetti2023proambition}. We have collaboratively developed a laboratory environment with various sensors and medical props (cf.~Fig.~\ref{fig:iot-setup}) to simulate and monitor the execution of treatment processes with medical students and trainees as target group. The focus hereby lies on hand hygiene as a process activity because this is an essential part of almost all medical procedures, but also an activity which is often neglected in practice~\citep{WHO2023_hand_hygiene_facts}. The group of medical experts comprised: two medical doctors, a nurse, a medical doctor in a management position (head of Infectiology), and two medical educators instructing students in our medical master program. We conducted four half-day workshops over two years (2023--2024) with these medical professionals serving as expert group to discuss and verify the use case, BP, sustainability-related aspects and indicators, and IoT aspects.

\subsection{Step 1: Identify sustainability dimensions and values}

\begin{figure}
    \centering
    \includegraphics[width=1.0\linewidth]{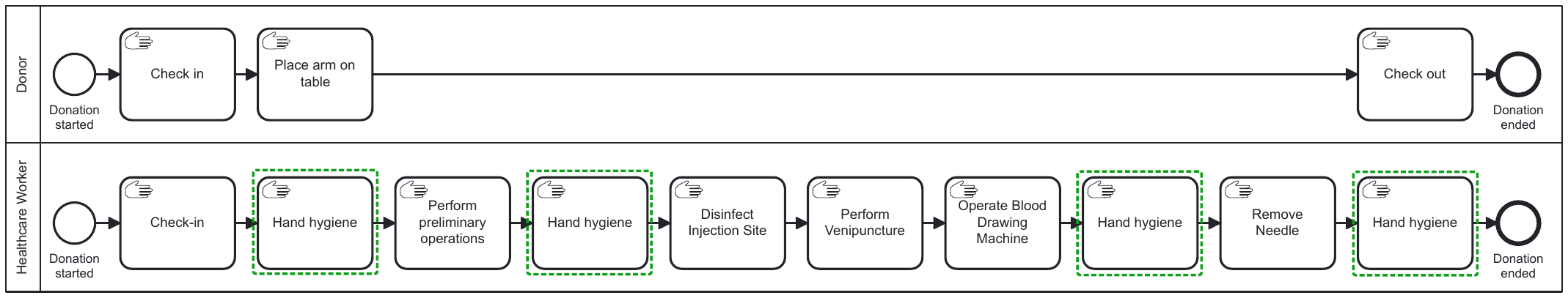}
    \caption{Established blood donation process (Task 1.1) with highlighted process fragments relevant for sustainability (Task 2.1).}
    \label{fig:bp-example}
\end{figure}

\begin{enumerate}
    \item Task: \emph{Establish the BP}. Adhering to hand hygiene is one of the most important and effective means to contain the spread of infections. The hand hygiene guidelines published by the World Health Organization (WHO)~\citep{10665-44102} serve as foundation for us to establish a specific medical procedure as a BP. These guidelines prescribe the points in time \emph{When} a hand hygiene has to be performed (e.g., before and after contact with patients, after touching contaminated surfaces), and they also recommend the amount of hand sanitizer (3--5 ml) to be used and the duration (approx.~30 seconds) of one execution of the hand hygiene activity. As a specific example of a medical procedure that includes multiple occurrences of the hand hygiene, we take the process of \emph{blood donation} (\emph{phlebotomy}), which should serve as the BP in this case study. Based on guidelines provided by the WHO~\citep{10665-44294}, we have modeled to specific process involving one donor and one healthcare worker (HCW) performing the procedure as process in BPMN~2.0 (cf.~Fig.~\ref{fig:bp-example}). The process involves a sequence of manual tasks to be performed by the HCW in our laboratory setup~\citep{franceschetti2023proambition}. The medical doctors and nurse acted as process owners and confirmed the validity of the modeled process. The dataset provided by~\cite{cruz2026labeleddatasetsimulatedphlebotomy} contains annotated images with objects and human-object annotations to further illustrate the phlebotomy process and setup.
    \item Task: \emph{Select Sustainability Dimensions Relevant for Evaluation and/or Improvement}. The main sustainability dimension is the individual (human) sustainability as the main cause of infection prevention is preserving human health. Along with the use of the hand sanitizer for hand hygiene, we are also interested in the environmental and economic sustainability dimensions as, on the hand, the sanitizer (alcohol-based chemicals) have an impact on the environment, and on the other hand, they are consumables that have to be purchased. The head of the department of Infectiology served as senior executive here in selecting these dimensions.
    \item Task: \emph{Select Values Related to Sustainability Dimensions}. The discussions with the medical experts--with the nurse in the lead as sustainability expert--revealed the following values related to the three dimensions: 1) compliance with hand hygiene standards and, 2) economical use of hand sanitizer. 
\end{enumerate}

\subsection{Step 2: Measure sustainability of the Business Process}

\begin{enumerate}
    \item Task: \emph{Identify Process Fragments Relevant to Selected Values}. The process fragments relevant to the selected values are all activities labeled \emph{Hand hygiene} in the BP model (cf.~Fig.~\ref{fig:bp-example}). These prescribe points in the normative process where a hand hygiene has to be performed. The sustainability expert confirmed this selection.
    \item Task: \emph{Propose Indicators for Process Fragments}. For one execution of the BP (a \emph{case}), the \emph{number of hand hygiene executions and their points in time} can be good indicators for hand hygiene compliance (i.e., the HCW did not forget to perform a hand hygiene when it is required by the WHO guidelines). To allow for a complete analysis of process conformance, all activities performed by the HCW have to be tracked to determine full compliance. For each fragment (i.e., hand hygiene activity), its \emph{duration} (recommended at least 30 seconds); and the \emph{amount} of alcohol-based sanitizer used (recommended 3--5 ml) are indicators for hand hygiene compliance and the economical use of sanitizer. With all three values, the general rule \emph{the more, the merrier} is applicable. However, performing the hand hygiene too often, for too long, and using too much sanitizer might have negative impacts on the environmental and economical dimensions.    
    \item Task: \emph{Define Measurements for Indicators}. The point in time of performing the hand hygiene activity can be derived through the use of the hand sanitizer bottle, the amount of sanitizer based on the difference of the amount before and after the hand hygiene activity, and the duration based on the time difference between start and end of the activity.
    \item Task: \emph{Recommend IoT Devices for Monitoring of Indicators}. From our experiments working with IoT devices and sensors, we propose to install an IoT-enabled scale to monitor the use and amount of hand sanitizer from a bottle placed on the scale, and additional sensors (motion sensor, distance sensor) to monitor the presence of the HCW (cf.~Fig.~\ref{fig:iot-data}, left). Additionally, we propose to use buttons to manually track the hand hygiene activities for a more robust detection of this activity and its duration.
    \item Task: \emph{Install and Configure IoT Devices}. In our setup, we deploy the scale, motion and distance sensor, and the button(s) from the \emph{TinkerForge} vendor. These sensors are wired to a \emph{Master Brick}, which is connected via USB to a desktop computer. An application written in Python uses the sensor API to collect the data when sensor data changes, adds a timestamp, and writes it as a time series in an InfluxDB database. An example representation of the sensor data in JSON format is depicted in Fig.~\ref{fig:iot-data} (bottom left), and a visualization of the timeseries data in a Grafana dashboard can be found in Fig.~\ref{fig:iot-data} (right).
    \item Task: \emph{Recommend Business Process Data for Measurement of Indicators.} In addition to the IoT data, the process owner recommends the tracking of the activity executions (i.e., their start and end) as specified in the BPMN model describing the BP. This involves using a BPM system (e.g., Camunda) to track the execution of the activities as manual tasks and create an event log containing timestamped process events in XES format. 
\end{enumerate}

\subsection{Step 3: Analyze collected sustainability data}

\begin{figure}
    \centering
    \includegraphics[width=1.0\linewidth]{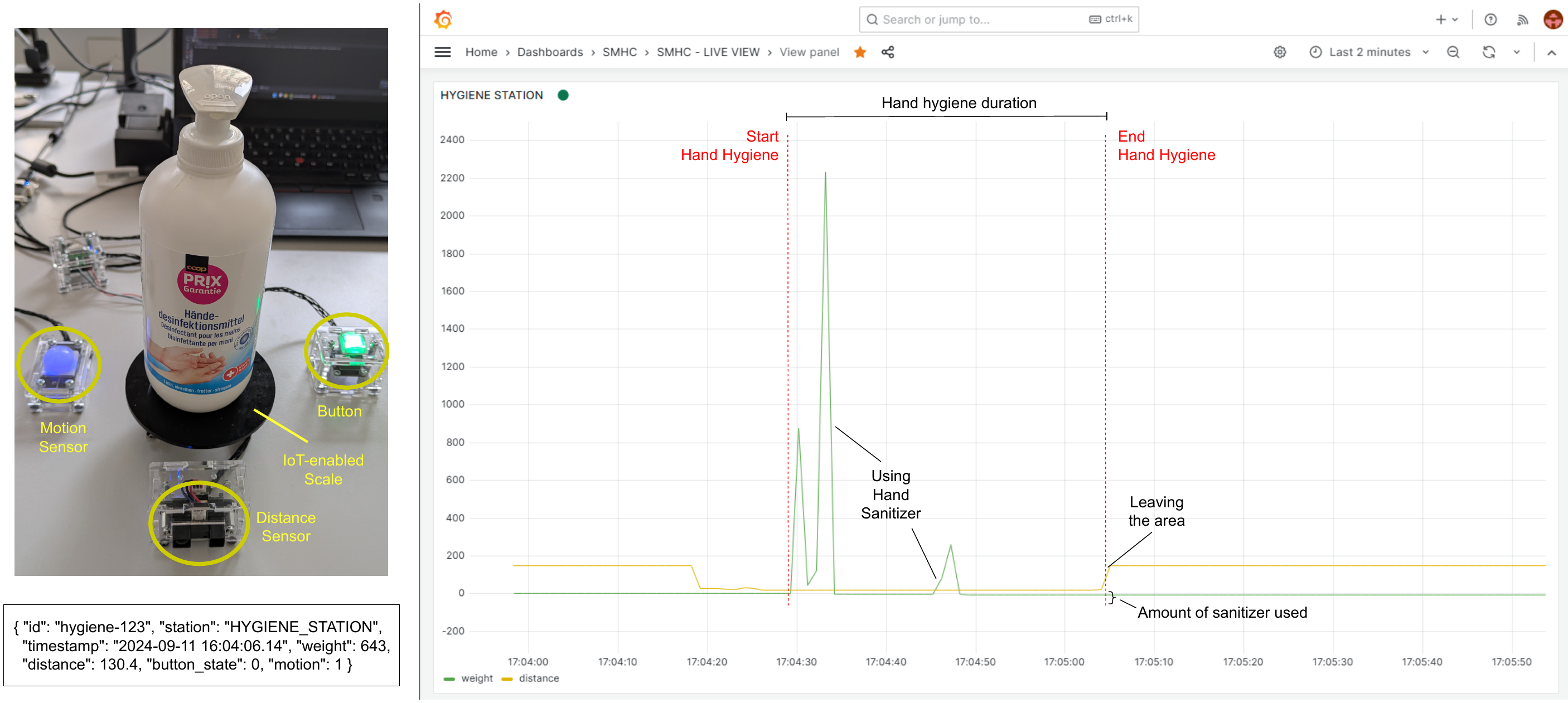}
    \caption{Collected IoT data from sensors (Task 3.1) and their analysis (Task 3.2).}
    \label{fig:iot-data}
\end{figure}

\begin{figure}
    \centering
    \frame{\includegraphics[width=1.0\linewidth]{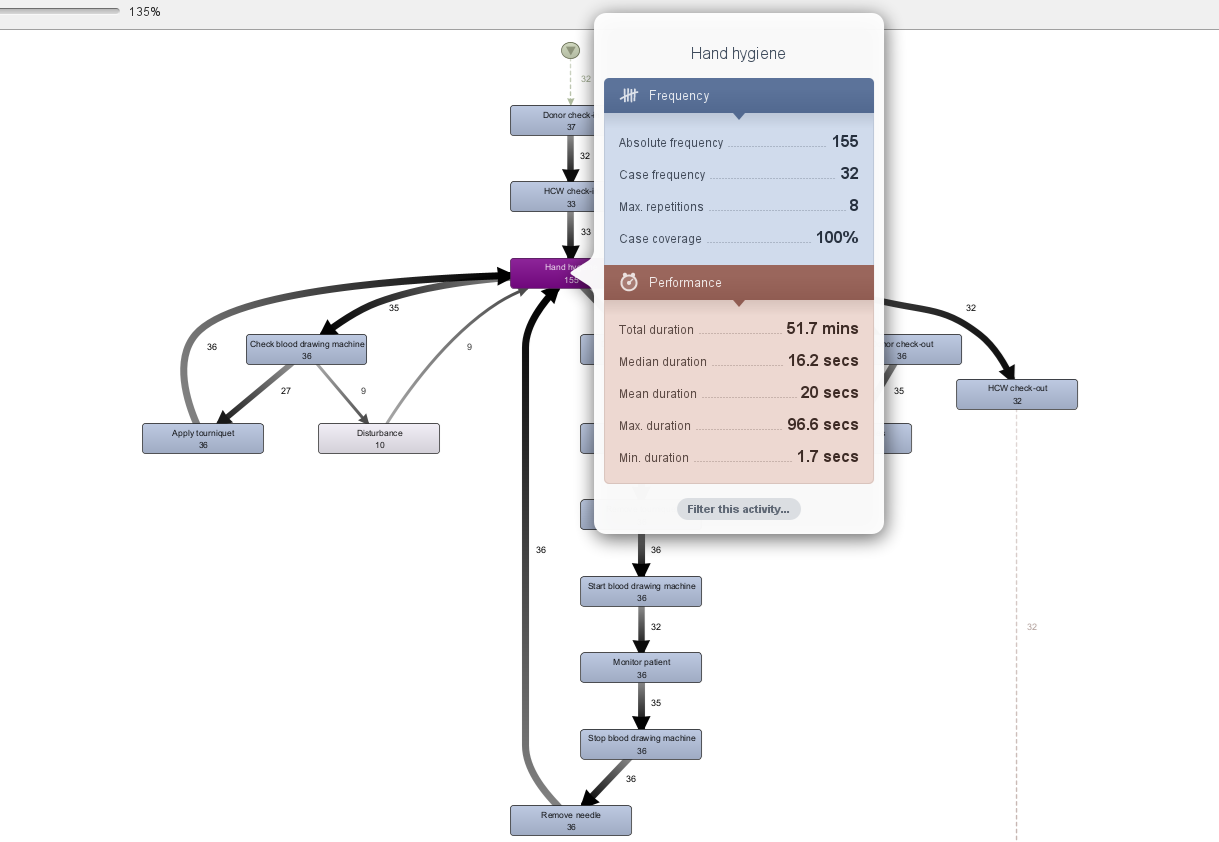}}
    \caption{Hand hygiene activity and process statistics derived from BP data (Task 3.2).}
    \label{fig:disco}
\end{figure}

\begin{figure}
    \centering
    \includegraphics[width=1.0\linewidth]{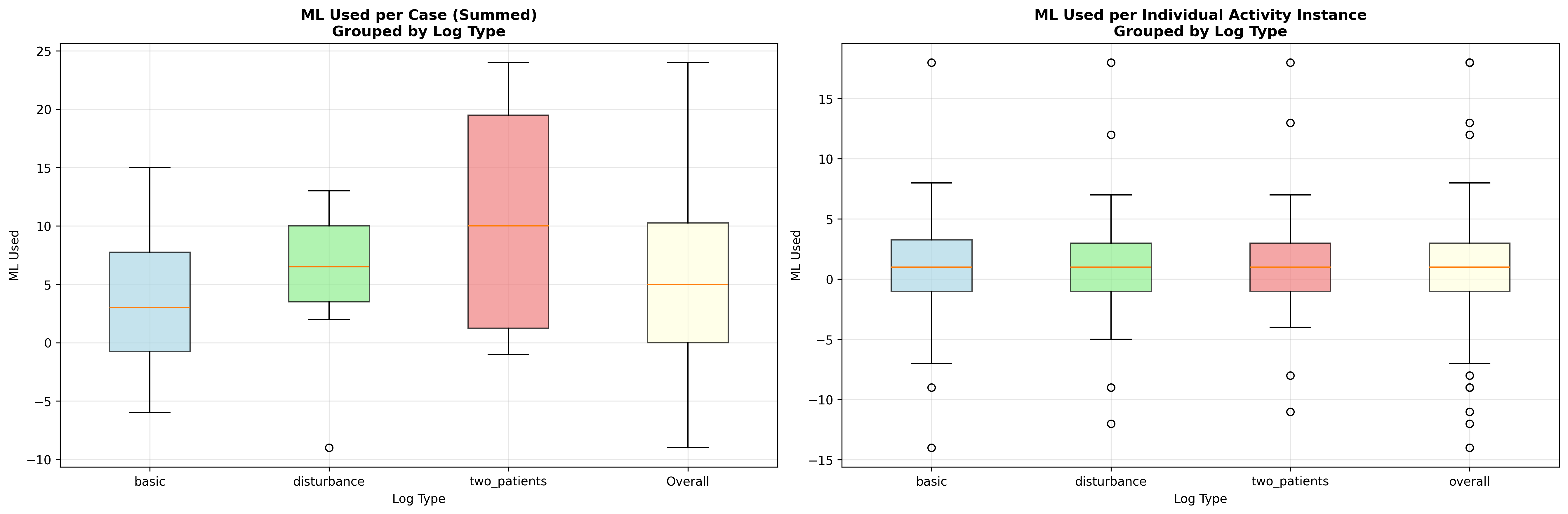}
    \caption{Hand hygiene statistics (used amount of sanitizer for different process scenarios) per case (left) and per activity (right) derived from IoT data (Task 3.2).}
    \label{fig:iot-statistics}
\end{figure}

\begin{enumerate}
    \item Task: \emph{Collect Measurement Data.} Using the IoT setup described in Task~2.5 and BPM system described in Task~2.6, we conducted two consecutive half-day workshops, corresponding to the observation period, with in total 17 first-year students of our medical master program in Summer 2024. The students were tasked with execution of the \emph{basic} blood donation process as described in Task~1.1 in our laboratory. This process model was displayed on a screen during the experiments such that the students can follow and consult the normative sequence of steps. While they were executing the BPs, the IoT monitoring system recorded the sensor data in the database, and we had an external supervisor monitoring the execution of the activities and creating a process event log for the Camunda BPMS. The dataset can be found in~\cite{seiger_2025_16736845}. To add more variations to the BP, we introduced some occasional \emph{disturbance} (e.g.,~a simulated phone call) to distract the students and trigger the need of an additional hand hygiene step after contaminating their hands. Furthermore, we simulated several instances of \emph{two patients} being treated by one HCW simultaneously. To complement the data collected automatically during the process executions, we collected qualitative data from the students as frontline personnel via a questionnaire regarding the usefulness and sustainability--loosely related to the defined sustainability indicators--of the BP and the IoT-based monitoring in our lab setup. These also included questions regarding the potential improvement of the BP and end-user feedback to be provided by a system executing an IoT-enhanced version of this process.
    \item Task: \emph{Derive Sustainability Indicators from Data}. The data engineer analyzed the IoT data and event log data (BP data) collected over the observation period of two days to calculate the sustainability-related values as published in~\cite{seiger_2025_16736845}. The data from the scale was used to identify the start of a hand hygiene activity (first occurrence of a peak in the scale readings) and the distance sensor was used to identify its end assuming that the hygiene area has been left by the HCW (cf.~Fig.~\ref{fig:iot-data}). In~\cite{seiger2025online}, we describe a more sophisticated approach to derive process activity executions from IoT data streams. This data were used in combination with the manual tracking of activity executions to create a process event log in XES format. From this data, typical process statistics including the process flow, number of cases, the number of hand hygiene activities and their durations (min, max, mean, median, etc.) were calculated using a process mining tool (cf.~Fig.~\ref{fig:disco}). Based on the identified start and end of a hand hygiene activity, we calculated the amount of used hand sanitizer for an activity from the IoT data. This calculation is based on the difference in the weight before the start and after the end of an instance of this activity, while filtering out the peaks that that occur during the activity because of the pressure on the sanitizer bottle when used (cf.~Fig.~\ref{fig:iot-data}). The summarized hand hygiene statistics including the amount of used sanitizer in milliliters derived from the IoT data are depicted in Fig.~\ref{fig:iot-statistics} per case (left) and per activity instance (right), each categorized by the scenario process (i.e., the basic blood donation process, the process with disturbance, and the process with two patients). Note that in a small number of instances, the measured amounts of used hand sanitizer were negative due to some calibration issues with the scale. 
    \item Task: \emph{Analyze Sustainability of BP Fragments}. The sustainability expert analyzed the derived sustainability indicators and associated process fragments. The statistics in Fig.~\ref{fig:disco} show that the mean duration of 20 seconds for the hand hygiene are below the official recommendations by the WHO (30 seconds), which indicates room for improvement. A more detailed analysis of the dataset revealed the strict conformance with the hand hygiene regulations according to \emph{When} it has to be executed. Here, we did not observe any deviations, which is highly relevant for preventing the spread of infections and maintaining the individual (human) sustainability. Note that the supervised training setting in our lab and the availability of the process model displayed on a screen facilitated this high level of hand hygiene adherence. In more realistic and noisy hospital settings, these statistics are likely to be worse~\citep{WHO2023_hand_hygiene_facts}. Furthermore, Fig.~\ref{fig:iot-statistics} (right) shows that the average amount of hand sanitizer used per activity is at the lower end of the recommendations (3--5 ml), but within range, which also means that there has been no observation of excessive use of hand sanitizer. All these insights have been compiled into an extensive sustainability report. 
    
    The results of the interview study with the medical students were also processed and added to this report. In summary, the students highlighted this kind of IoT-based monitoring system as being very useful in training settings as the monitoring setup reminded them to closely pay attention to the correct steps and hand hygiene activities. They see high potential of this setup to enhance the sustainability of the treatment processes using IoT by providing live feedback about the correct execution of the activity sequences and usage of hand sanitizer, especially in more noisy and realistic hospital settings. On the negative side, they acknowledged the increased technical effort related to the sensor setup~\citep{albert2024sustainability} and the more invasive monitoring of their actions. Regarding live feedback related to their performance of the hand hygiene activities, the students highlighted the need for unobtrusive, preferably haptic or visual means informing them about the current hygiene-related statistics and process conformance.
\end{enumerate}

\subsection{Step 4: Improve the Business Process}

\begin{figure}
    \centering
    \frame{\includegraphics[width=0.75\linewidth]{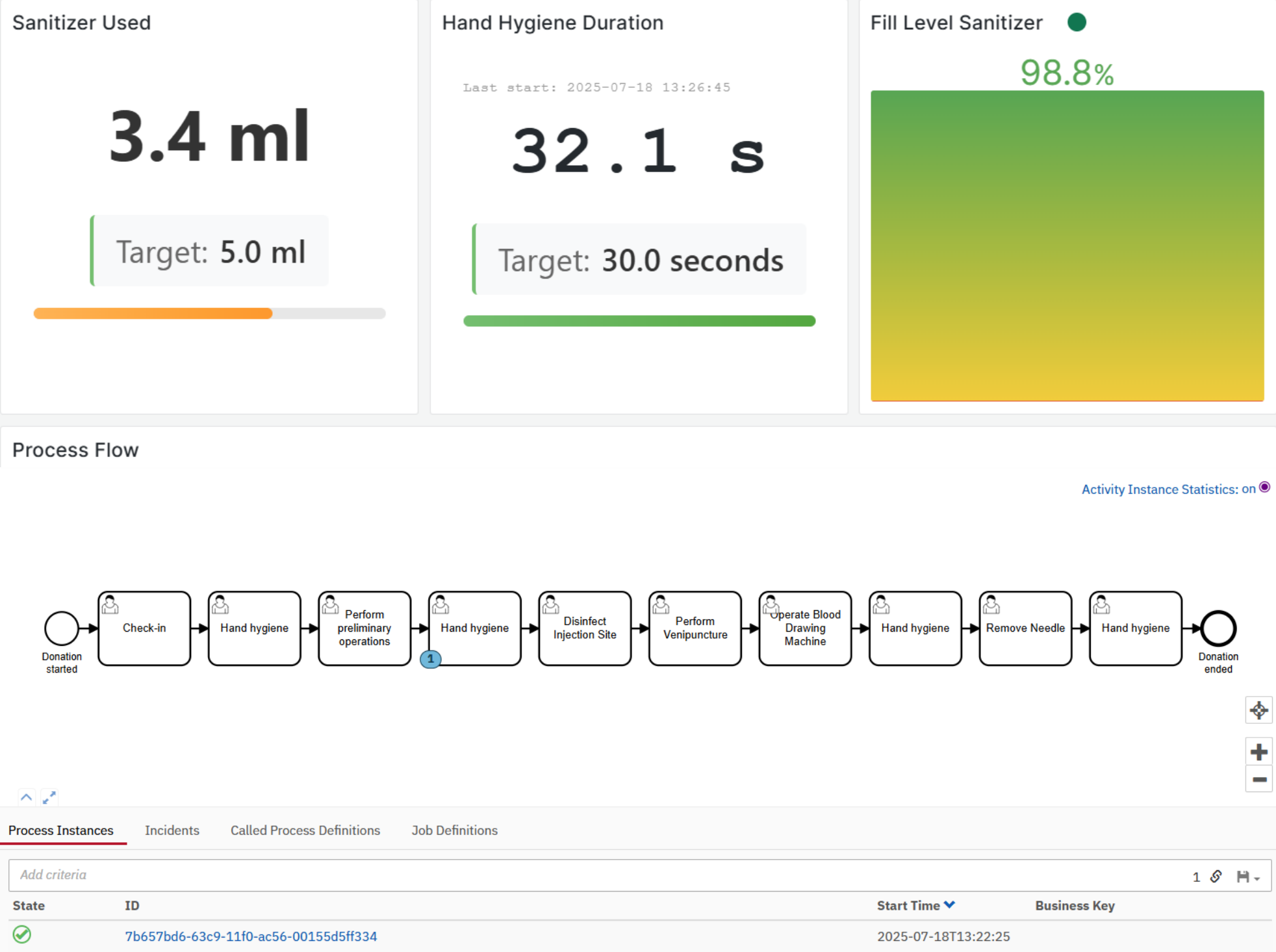}}
    \caption{Enhanced hand hygiene and process visualization for end-users in a dashboard as part of the process improvements (Task 4.2).}
    \label{fig:iot-improved}
\end{figure}

\begin{enumerate}
    \item Task: \emph{Propose Sustainable Activities}. Based on the report and insights from Task~3.3, the sustainability expert--the nurse in our team of experts--suggested two sustainable activities: 1)~to keep the sensor setup initially used for measuring the sustainability of the blood donation process and detecting activities, and use it within future training sessions for medical students. Considering the students' feedback from the surveys, the expert suggested: 2)~to extend the system with live visual feedback including the normative process model and current step, as well as statistics about the current execution of a hand hygiene activity with current amount of sanitizer used and target amount, current duration and target duration, and current fill-level of the sanitizer bottle to show when it may need to be refilled. All this data can be derived using the sensor setup (cf.~Fig.~\ref{fig:iot-setup}). The BP itself is based on the guidelines for phlebotomy (blood drawing) drafted by the WHO~\citep{10665-44294}. There is no need to change it or enhance it with IoT devices as all the activities in the process have to be performed manually. However, process elements (e.g., \emph{events}) representing warnings could be integrated into the existing BP model to notify the HCWs about forgotten hand hygiene activities. Furthermore, the use of smart hand sanitizers with automatic dispense and counting of a specified amount of liquid could be considered in future setups. 
    \item Task: \emph{Select Sustainable Activities}. The process owners--the medical doctors--reviewed suggestions~1) and~2) and deemed them suitable and feasible to be implemented as they are of low effort. The sensors, data collection, and data processing infrastructures already exist, only the visualizations have to be developed. The extension of the BP model with further events and integrating the corresponding conformance checking mechanisms to detect deviations will be considered in future iterations.
    \item Task: \emph{Translate Actions into Business Process Modifications}. The BP modifications are minimal to adapt the sustainable actions, they only consider the infrastructure supporting the BP executions and enhancing it with IoT. The modifications comprise: 1)~fixing the sensor infrastructure to monitor activity executions (scale, distance and motion sensor) in a fixed setup; and 2)~installing and extending the data collection and visualization infrastructure. We extended our existing Grafana-based dashboards with new visualizations that show the current amount of sanitizer used, current duration of hand hygiene, and the current fill-level of the sanitizer (cf.~Fig.~\ref{fig:iot-improved}, top), as well as the visualization of the current process progress based on the Camunda dashboard visualizations (cf.~Fig.~\ref{fig:iot-improved}, bottom).
    \item Task: \emph{Validate and Finalize Business Process Updates}. The senior executive--head of Infectiology--reviewed these changes and acknowledged their correctness and appropriateness. In a follow-up study with the next cohort of medical students, we will use the IoT-based monitoring infrastructure and extended visual feedback to monitor the hand hygiene executions and evaluate the students' adherence. Especially with presenting the current status and target values for both, the amount of sanitizer used and hand hygiene duration, we expect the students to have a better awareness of these two sustainability-related values and thus, hand hygiene statistics to improve and become more sustainable. 
\end{enumerate}

\section{Discussion} \label{sec:discussion}

This section revisits the research question (RQ) posed at the beginning of the article (cf.~Section~\ref{sec:problemStatement}): ``How can we provide methodological support to (1)~analyze and, (2)~improve the sustainability of Business Processes by using IoT?''. It elaborates on the key aspects that make this methodological support both systematic and adaptable to real-world organizational contexts. First, we discuss our findings from a conceptual analysis and insights from the controlled case studies, addressing specific criteria related to the applied design science research (DSR) method based on~\cite{hevner2004design,hevner2007three,peffers2007design}. We then elaborate on further relevant aspects and practical limitations.

\paragraph{DSR Criteria}

\begin{itemize}
    \item \textbf{Functionality/Utility}: The two developed artifacts--the conceptual model and methodology--address a relevant real-world problem, namely the evaluation and improvement of the sustainability of business processes by means of IoT. The methodology provides structured guidance for sustainability analysis and for identifying improvement opportunities, specifically in contexts where BP are sufficiently instrumented and IoT data is available. We illustrated the utility of both artifacts to be useful in a realistic context in two exemplary, illustrative case studies from two different domains, providing initial evidence of applicability across contexts. Larger-scale deployments remain subject to future work.  
    \item \textbf{Effectiveness}: We demonstrated the model and method's effectiveness based on the tourism example and smart healthcare case study. The application in these examples provides initial evidence of the plausibility of our proposed artifacts, showing that the approach can be applied in realistic settings and can generate meaningful insights. However, we did not perform a quantitative comparison of the processes before and after sustainability improvements and also no comparison with alternative methods. These will be considered in future work to provide stronger empirical evidence of the effectiveness of the proposed artifacts.
    \item \textbf{Efficiency and Reliability}: A limitation of the presented example and case study is that we do not assess, nor indicate, the time and cost overhead as well as the computational and organizational effort of applying the proposed methodology. The method involves several tasks that require detailed data collection, analysis, trade-off discussion, and decision-making by different stakeholders. Furthermore, the provisioning, installation, programming, and analysis of data from the IoT devices introduce considerable costs and technical challenges. Therefore, conducting systematic studies of additional real-world use cases involving external users applying the methodology is required to provide more insights regarding the methodology's efficiency. Furthermore, consistency of results has to be demonstrated in future case studies to show empirically that the methodology produces reliable outcomes. 
    \item \textbf{Usability}: The methodology is presented as a structured set of tasks described with a description, required inputs, expected outputs, and involved roles. The graphical BPMN~2.0 models illustrate the specific activity flows and input/output artifacts using a standard notation for BPs. Note that BPMN provides an intuitive and easy way to represent the semantics of complex processes. This notation is not only used by process designers who are experts in it, but also by other process stakeholders such as customers, marketing professionals, or finance employees who simply need to analyze it \citep{Harmon2011BPMsurvey,Leopold2016BPMNquality,Nysetvold2005QualityFramework}. In addition, to obtain initial feedback, we presented the methodology and conceptual model to an expert from the smart tourism domain and discussed them informally. Even though the expert's feedback indicated a positive perception of usability, more comprehensive user studies to evaluate the usability of the proposed artifacts are required to provide a more rigorous basis for evaluation. 
    \item \textbf{Scalability and Robustness}: In this article we illustrate the conceptual model and methodology based on a small running example from smart tourism and a more comprehensive controlled case study experiment in smart healthcare. While both use cases are rather small in scope and complexity and driven by us, larger scale validations in enterprise contexts and different domains with BPs of different complexity levels have to be considered in future work.
    \item \textbf{Generalizability}: Both the conceptual model and the methodology are designed to analyze and improve sustainability of business process using IoT. To this end, all the relevant concepts and steps are integrated in a domain-agnostic way, which is one of the strengths of conceptual modeling. We demonstrated two instantiations in smart tourism and smart healthcare to illustrate applicability across domains. Further instantiations and studies are required to support the claims of generalizability. We expect that the conceptual model and methodology can also be applied to BPs in other IoT-related domains, e.g., smart homes and offices, and smart manufacturing.
    \item \textbf{Novelty and Innovation}: The sustainability-related considerations in our investigations go beyond the ``traditional'' environmental aspects, representing an advancement over related research. While several approaches exist that address and combine sustainability-related research with BPM research (cf.~Section~\ref{sec:related}), we add the new perspective of IoT as an enabler and potential solution to improve the sustainability of BPs. IoT devices provide advanced sensing and automation capabilities, with a high potential to improve the sustainability of BPs along several dimensions across many different domains. Our work provides a novel and clear conceptual and methodological integration of these three different research fields.
    \item \textbf{Relevance and Contribution}: Our developed artifacts--the conceptual model and the structured methodology--align well with the DSR knowledge contributions including design principles and conceptual frameworks. We consider these proposals as strong academic contributions of high theoretical relevance, novelty and topicality. With smart tourism and healthcare we have identified two specific domains that involve many manual tasks that can be supported and automated via IoT to improve their sustainability. Even though this indicates practical relevance, more real-world industrial deployments are required to validate the practical relevance of our work.
\end{itemize}

\paragraph{Traceability} The proposed conceptual model demonstrates how structured modeling, indicator-based assessment, and targeted IoT integration can enable both the evaluation (RQ--1) and the enhancement (RQ--2) of sustainability across multiple dimensions. A key strength of the methodology lies in its systematic, process-based nature, which enables documenting and tracing each step of the process, including the relevant stakeholders, and the required inputs and outputs. This traceability is particularly relevant when addressing sustainability goals, as it ensures that all stakeholders can understand why certain choices were made, what trade-offs were considered, and how the process evolved in response to sustainability indicators or IoT-enabled insights. Such transparency not only improves the quality of decision making, but also facilitates continuous improvement through reflective practice.

\paragraph{Conceptual clarity}
The methodology also contributes conceptually by clearly defining key notions from the Sustainability, BPM, and IoT domains, and explicitly modeling their interrelations. This integrated view enables the formulation of tailored sustainability metrics for IoT-enhanced BPs, grounded in the five presented sustainability dimensions. Such conceptual clarity is crucial not only for implementation, but also for benchmarking, comparison across processes, and long-term tracking of sustainability improvements.

\paragraph{Automation potential}
The explicit modeling of BP steps allows organizations to analyze and optimize their workflows, revealing inefficiencies, redundancies, or sustainability bottlenecks. More importantly, this formal representation facilitates the identification of automation opportunities, particularly when supported by Model-Driven Development (MDD) techniques and IoT devices. More specifically, when IoT devices are integrated into these processes, we are able not only to collect real-time data for sustainability indicators, but also to trigger automated actions that can contribute to improvements across the various sustainability dimensions that an organization aims to address (e.g., reducing waste, improving energy efficiency, or enhance the user experience)~\citep{albert2024sustainability,DBLP:journals/jss/ValderasTS22}. Therefore, the methodological rigor of process modeling ensures that such automation is both context-aware and sustainability-aligned.

\paragraph{Clear responsibilities}
Another essential component is the explicit identification of the actors responsible for each process step. This visibility supports the need to conform and coordinate interdisciplinary teams, bringing together experts from business process management (BPM), sustainability science, IoT engineering, and potentially data science or human factors. Understanding how these actors interact and collaborate across disciplinary boundaries fosters a more holistic design approach and encourages shared ownership of sustainability outcomes, which are often multifaceted and cross-cutting in nature. In this context, we see high potential for using Large Language Models (LLMs) to assist in various stages of the proposed methodology. We envision that well-trained LLMs can assist stakeholders along the methodology steps for example to propose indicators to evaluate and measure process fragments regarding specific values, suggest candidate IoT devices to measure and/or automate existing tasks, and propose changes in current processes to align them with specific sustainability goals.

\paragraph{Trade-off Management}

In several parts of the methodology, both the sustainability expert and the IoT expert are challenged by potentially complex dependencies among sustainability dimensions and associated values. When selecting sustainability dimensions and associated values to be measured (Task~1.3) in order to identify potential improvements, the sustainability expert must already consider trade-offs and possible negative impacts on other dimensions. Similarly, in Task~2.4, the IoT expert needs to assess the sustainability implications of installing and operating new IoT devices for monitoring purposes. When the sustainability expert suggests business process improvements (Task~4.1), these changes must be evaluated concerning their impact on other sustainability dimensions that are not the primary focus of the improvements. Managing these trade-offs is a complex task that requires domain expertise, as well as careful analysis and prioritization in alignment with business goals and requirements. Decisions should be guided by the relative importance of the affected sustainability dimensions, organizational priorities, feasibility, and expected return on investment, ensuring that trade-offs are addressed systematically and transparently rather than implicitly. At the same time, the current methodology does not provide explicit mechanisms or formal decision-support techniques to structure this process. Given the complexity involved, a more comprehensive treatment of trade-off management is beyond the scope of this work. As part of future research, we plan to extend the methodology with a formal decision-support layer, for example, by incorporating weighted indicators or Multi-Criteria Decision Analysis, to support the ranking of alternatives when improvements in one sustainability dimension negatively affect others.

\paragraph{Practical Limitations}

In addition to the limitations identified in the DSR criteria discussion, we see further practical challenges and limitations of the proposed methodology: 1)~The methodology is relying heavily on different types of data (generated from IoT, from process executions, and from metrics related to sustainability). Organizations may not have sufficient infrastructure to collect all the data; data might be incomplete, noisy, and not integrated across systems. 2)~Organizations typically have legacy IT systems and heterogeneous process landscapes. The integration of IoT devices, process monitoring tools, and sustainability indicators might be technically and organizationally challenging. 3)~As discussed, sustainability is multi-dimensional, and often hard to quantify. Organizations must define specific KPIs and measurement methods, which is not always straightforward. 4)~The effective application of the methodology requires expertise and interdisciplinary knowledge in the areas of BPM, IoT, and sustainability. Multiple experts have to be involved to carry out the method. The extensive analysis tasks might introduce considerable time and cost overheads, which we have not yet analyzed. There is a uncertainty regarding the cost-benefit  of applying the methodology that needs to be clarified to estimate whether and under which conditions there is a return on investment. 5)~So far, there is limited tool support to operationalize the application of the methodology. We are currently working on a first set of tools and extensions to support users.  
\\
\\
\noindent
In summary, the proposed methodology offers a structured and comprehensive response to the considered research question by combining conceptual clarity, process modeling rigor, and practical integration of IoT technologies. Its systematic approach facilitates traceability, transparency, and continuous improvement, while the explicit representation of process elements supports both optimization and automation aligned with sustainability goals. The clear identification of involved actors fosters interdisciplinary collaboration, and the use of domain-specific notions enables the formulation of relevant sustainability metrics across all five dimensions. Moreover, the potential integration of LLMs introduces an intelligent, adaptable layer that can further assist organizations in identifying IoT solutions tailored to specific sustainability challenges. Altogether, these elements demonstrate how methodological support can be effectively designed to evaluate and enhance the sustainability of IoT-enhanced BPs.

Our work improves the state of the art by introducing a systematic methodology that not only focuses on the improvement of sustainability in BPM but also considers IoT infrastructures as a key factor in this improvement. In particular, the proposed methodology supports the improvement of sustainability in IoT-enhanced BPs by precisely defining the steps to be done in this type of processes, identifying the actors that must perform each step, and proposing the inputs and outputs of these steps. This provides a valuable mechanism for guiding the analysis of sustainability and supporting more informed decisions about improving it within IoT-enhanced BPs. In addition, our methodology does not focus only on the environmental dimension as most of the analyzed related works do (cf.~Section~\ref{sec:related}), but it provides a holistic solution considering the five dimensions identified by the existing literature (environmental, economic, human, social, and technical).

\section{Conclusions and future work} \label{sec:conclusions}
This paper proposes a methodology for systematically evaluating and improving IoT-enhanced BPs with respect to sustainability. Addressing a gap in the business process domain, where sustainability is often approached in an ad hoc manner, the methodology provides business process owners with a tool to identify specific activities within their business processes that can be improved to generate various sustainability-related benefits. It defines a set of steps, each with its specified inputs, outputs, its main actors required for execution, and illustrated with an example thus facilitating its application.

The process begins with the identification of sustainability dimensions that are relevant to the business and concludes with the measurement of sustainability indicators related to those dimensions. It then proposes modifications to the BPM, such as the inclusion or adaptation of activities, aimed at improving sustainability performance in the targeted dimensions. The approach leverages IoT devices, which serve as sensors to measure sustainability indicators and as actuators to enhance sustainability aspects within specific BPM activities. The methodology is grounded in a conceptual model that defines the core concepts shared across the domains of sustainability, business processes, and IoT technologies. This conceptual model builds upon previous models of sustainability and business processes. To validate the approach, the paper presents a case study in the healthcare domain where the proposed methodology is applied. The case illustrates how the methodology can be used to identify relevant sustainability dimensions, measure the corresponding indicators through IoT devices, and propose a process modification to improve sustainability performance. This application demonstrates the feasibility and value of the methodology.

Future work will focus on several directions. First, we plan to advance our ongoing efforts to incorporate LLMs as actors within the methodology to (partially) automate tasks currently performed by sustainability and IoT experts. We are currently developing two task-specific LLMs: one aimed at supporting sustainability-related decisions (e.g., selecting values, indicators, and measurements); and another LLM focused on recommending IoT devices to support sustainability actions and indicator monitoring. These models are trained on datasets that integrate standards, regulations, process patterns, and IoT device specifications--enabling a data-driven approach to enhance the methodology's usability. Second, we plan to conduct additional case studies across different real-world domains, involving end users to assess the methodology's applicability. Finally, we aim to carry out multiple iterations of the methodology within the same process context, enabling the analysis of the impact of proposed modifications and supporting continuous improvement in sustainability performance.

\bmhead{Acknowledgements}
This work has received funding from the Swiss National Science Foundation under Grant No.~IZSTZ0\_208497 (\textit{Pro\-AmbitIon} project) and Grant No.~10002384 (\emph{TaSSAreCt} project), from the Research Vice-Rectorate of the Polytechnic University of Valencia (UPV) with the Aid to First Research Projects (PAID-06-22), and from the MICIU/AEI/10.13039/501100011033, FEDER, and the UE under Grant No.~PID2023-146224OB-I00 (\textit{PRODIGIOUS Project}).

\bibliography{references}

\end{document}